\documentclass[longauth]{aa}
\usepackage[T1]{fontenc}

\DeclareRobustCommand{\VAN}[3]{#2}
\let\VANthebibliography\thebibliography
\def\thebibliography{\DeclareRobustCommand{\VAN}[3]{##3}\VANthebibliography}

\usepackage{graphicx}	
\usepackage{amsmath}	
\usepackage{amssymb}	
\usepackage{ulem}
\usepackage{natbib}
\usepackage{hyperref}

\usepackage{caption}
\usepackage{subcaption}
\usepackage{color}

\setcitestyle{citesep={,}}

\newcommand{\Tab}[1]{Table~\ref{#1}}

\newcommand{\lla}{\hbox{$\lambda\lambda 2796,2803$}}
\newcommand{\Wr}{\hbox{$W_{\rm r}^{2796}$}}

\newcommand{\CIV}{\hbox{{\rm C}{\sc \,iv}}}

\newcommand{\FeII}{\hbox{{\rm Fe}{\sc \,ii}}}

\newcommand{\OII}{\hbox{[{\rm O}{\sc \,ii}]}}

\newcommand{\OIII}{\hbox{[{\rm O}{\sc \,iii}]}}
\newcommand{\MnII}{\hbox{{\rm Mn}{\sc \,ii}}}
\newcommand{\MgI}{\hbox{{\rm Mg}{\sc \,i}}}
\newcommand{\MgII}{\hbox{{\rm Mg}{\sc \,ii}}}
\newcommand{\CaII}{\hbox{{\rm Ca}{\sc \,ii}}}
\newcommand{\NII}{\hbox{[{\rm N}{\sc \,ii}]}}
\newcommand{\SII}{\hbox{[{\rm S}{\sc \,ii}]}}
\newcommand{\NeIII}{\hbox{{\rm Ne}{\sc \,iii}}}

\newcommand{\HI}{\ion{H}{i}}

\newcommand{\lya}{\hbox{{\rm Ly}$\alpha$}}
\newcommand{\Ly}{\hbox{{\rm Ly}$\alpha$}}
\newcommand{\Ha}{\hbox{{\rm H}$\alpha$}}
\newcommand{\Hb}{\hbox{{\rm H}$\beta$}}
\newcommand{\Hg}{\hbox{{\rm H}$\gamma$}}
\newcommand{\Hd}{\hbox{{\rm H}$\delta$}}

\newcommand{\fdlam}{erg\,s$^{-1}$\,cm$^{-2}$\,\AA$^{-1}$}

\newcommand{\flux}{erg\,s$^{-1}$\,cm$^{-2}$}

\newcommand{\mpy}{\hbox{M$_{\odot}$\,yr$^{-1}$}}

\newcommand{\msun}{\hbox{M$_{\odot}$}}

\newcommand{\kms}{\hbox{km~s$^{-1}$}}
\newcommand{\kpc}{\hbox{kpc}}
\newcommand{\mAB}{\hbox{m$_{\rm AB}$}}
\newcommand{\LOII}{\hbox{L}$_{\OII}$}

\newcommand{\gpk}{{\textsc GalPaK$^{\rm 3D}$}}{}

\newcommand{\detLimOII}{3.7$^{+0.8}_{-0.6}\times10^{-18}$\flux}
\newcommand{\detLimCont}{26~\mAB}

\newcommand{\detLimOIIshallow}{7.07$^{+1.6}_{-1.3}\times10^{-18}$\flux}
\newcommand{\detLimContshallow}{25.2~\mAB}

\begin{document}


\title{MusE GAs FLOw and Wind (MEGAFLOW) XII. Rationale and design of a \MgII\ survey of the cool circum-galactic medium with MUSE and UVES: The MEGAFLOW Survey}


   \author{Nicolas  F.  Bouch\'e           \inst{1}
   \and
      Martin Wendt 	\inst{2}
 \and 
Johannes Zabl \inst{1,3}
\and 
Maxime Cherrey \inst{1}
\and
Ilane Schroetter \inst{4}
\and
Ivanna Langan \inst{1,5}
\and
Sowgat Muzahid\inst{6}
\and
Joop Schaye \inst{7}
\and
Benoît Epinat \inst{8,9}
\and
Lutz Wisotzki \inst{10}
\and
Thierry Contini \inst{4}
\and
Johan Richard \inst{1}
\and
Roland Bacon \inst{1}
\and
Peter M. Weilbacher \inst{10}
}
\institute{Univ. Lyon1, Ens de Lyon,  CNRS, Centre de Recherche Astrophysique de Lyon (CRAL) UMR5574, F-69230 Saint-Genis-Laval, France\\
   \email{nicolas.bouche@univ-lyon1.fr}
\thanks{thanks}
\and Institut für Physik und Astronomie, Universität Potsdam, Karl-Liebknecht-Str. 24/25, 14476 Potsdam, Germany\\
\and Institute for Computational Astrophysics and Department of Astronomy \&\ Physics, Saint Mary’s University, 923 Robie Street
Halifax, Nova Scotia B3H 3C3, Canada
\and  Institut de Recherche en Astrophysique et Plan\'etologie (IRAP), Universit\'e de Toulouse, CNRS, UPS, F-31400 Toulouse, France\\
\and European Southern Observatory (ESO), 
Karl-Schwarzschild-Str. 2, 85748 Garching b. München, Germany\\
\and  Inter-University Centre for Astronomy \& Astrophysics (IUCAA), Post Bag 04, Pune, India, 411007\\
\and Leiden Observatory, Leiden University, PO Box 9513, NL-2300 RA Leiden, the Netherlands\\
\and Aix Marseille Univ., CNRS, CNES, LAM, Marseille, France \\
\and Canada-France-Hawaii Telescope, 65-1238 Mamalahoa Highway, Kamuela, HI 96743, USA\\
\and Leibniz-Institut für Astrophysik Potsdam (AIP), An der Sternwarte 16, 14482 Potsdam, Germany
}

\date{Accepted October 24, 2024. Received June 12, 2024; in original form \today}


\abstract{
We present the design, rationale, properties and catalogs of the MusE Gas FLOw and Wind survey (MEGAFLOW), a survey of the cool gaseous halos of $z\simeq1.0$  galaxies using  low-ionization \MgII{} absorption systems.
The survey consists of 22 quasar fields selected from the Sloan Digital Sky Survey (SDSS)  having multiple ($\geq3$) strong \MgII{} absorption lines over the redshift range $0.3<z<1.5$.
Each  quasar was observed
with  the Multi-Unit Spectroscopic Explorer (MUSE)  and the Ultraviolet and Visual Echelle Spectrograph  (UVES), for a total of 85~hr and 63~hr, respectively.
The UVES data  resulted in 127 \MgII{} absorption lines over $0.25<z<1.6$, with  a median rest-frame equivalent width (REW) $3\sigma$ limit 
of $\approx 0.05$~\AA.
The MUSE data resulted in $\sim$2400 galaxies of which 1403 with redshift confidence \texttt{ZCONF}$>1$, i.e. more than 60 galaxies per arcmin$^{2}$. 
They were identified using a dual detection algorithm based on both continuum and emission line objects.
   The achieved  \OII\ 50\%\ completeness is \detLimOII{} (corresponding to SFR$>0.01$~\mpy{} at $z=1$) using realistic mock \OII{} emitters
and the 50\%\ completeness is $m_{F775W}\approx26$ AB magnitudes for continuum sources. 
We find that (i) the fraction of \OII{} emitting galaxies  which have no continuum is $\sim15$\%;
(ii) the success rate in identifying
at least one galaxy within 500~\kms{} and 100~\kpc{} is $\approx90$\%{} for \MgII{} absorptions with $\Wr\gtrsim0.5$~\AA{} ;
 (iii) the mean number of galaxies per \MgII{} absorption is  $2.9\pm1.6$ within the MUSE field-of-view;
(iv) of the 80 \MgII{} systems at $0.3<z<1.5$, 40 (20) have 1 (2) galaxies within 100~kpc, respectively;
(v) all but two host galaxies have stellar masses $M_\star>10^9$~\msun, and star-formation rates $>1$~\mpy.
} 


 \titlerunning{The MEGAFLOW survey}
    \authorrunning{Bouché et al.}

\maketitle

\section{Introduction}
   The circum-galactic medium (CGM) is the complex interface between the intergalactic medium and the galaxies themselves.
 Traditionally, the CGM describes the
  gas surrounding galaxies outside their disks or interstellar medium and inside their virial radii \citep{TumlinsonJ_17a}.
The CGM is the interface between the continuous fresh source of fuel coming from the IGM
and star-formation driven outflows, often referred to as the baryon cycle \citep{PerouxC_20b}.
 As a result, the CGM is expected to retain the kinematical (and/or enrichment) signatures of these processes.

  A significant amount of effort has been devoted to the study of galaxies
close to quasi-stellar object (QSO) sight-lines to study the interplay between outflows and accretion. The idea of such a medium goes back to the detection of cold gas clouds toward stars at high Galactic latitudes \citep{Spitzer=1956}.
Soon after the discovery of the first quasars, \cite{Bahcall=1969} proposed that most of the absorption lines observed in QSO (quasar) spectra were caused by gas in extended halos of normal galaxies.

The most commonly used metal absorption in quasar
spectra is the magnesium (\MgII) \lla{} doublet, which has been known to   trace 'cool'  ($T \sim10^4$~K) 
photoionised gas in and around galaxies since the studies of \citet{BergeronJ_91a,BergeronJ_92a,BergeronJ_94a,SteidelC_93a} and \citet{SteidelC_94a}.
In the vicinity of galaxies,  \MgII{} absorptions are expected
to occur in sightlines probing  either outflows \citep[as in][]{NestorD_11a} or accreting gas in extended gaseous disks \citep{FumagalliM_11a,PichonC_11a,KimmT_11a,ShenS_13a,DeFelippisD_21a}.
 Strong \MgII{} absorptions, with REW $\Wr>1$~\AA{} can also occur in groups \citep[e.g.][]{KacprzakG_10b,GauthierJR_13a,BielbyR_17a}
and weak \MgII{} absorptions ($\Wr<0.1$~\AA) in the outskirts of clusters \citep[e.g.][]{MishraS_22a}.

The detection of tens of thousands of \MgII{} absorbers with $\Wr\gtrsim0.5$~\AA{}
\citep[e.g.][]{NestorD_05a,ZhuG_13a} thanks to  the Sloan Digital Sky Survey \citep[SDSS,][]{YorkD_00a} have enabled large statistical analyses \citep[e.g.][]{BoucheN_06c,LundgrenB_09a,GauthierJR_09a}.
As noted in \citet{NestorD_05a}, the REW $\Wr$ distribution is a double exponential, where the transition occurs at around $\Wr\simeq0.5$~\AA{} indicating a transition between strong and weak \MgII{} systems possibly related to different physical mechanisms.
This is supported by the different redshift evolution of strong and weak \MgII{} systems \citep{NestorD_05a}.

However, connecting the kinematics of the gas inflow and ouflow processes to the kinematics of the host galaxy requires us to be able to identify the host efficiently.
 The advent of the Multi Unit Spectroscopic Explorer
 \citep[MUSE,][]{BaconR_10a} 
 on the Very Large Telescope (VLT) has revolutionised the study of the CGM thanks to its exquisite sensitivity and its field of view (FOV).
As a result, a number of MUSE-based CGM surveys have been developed over the past few years such as QSAGE \citep{Bielby_2019}, MUSEQuBES \citep{MuzahidS_20a}, CUBES \citep{ChenHW_20a}, MAGG \citet{LofthouseE_20a}, and MUSE-ALMA \citep{PerouxC_19a}.


 This paper describes the rationale, design and properties of the MusE GAs FLOw and Wind (MEGAFLOW) survey,  a \MgII-selected survey around 22 QSO fields.  
This paper is part of a series of 13 papers  \citep{SchroetterI_16a,SchroetterI_19a,SchroetterI_21a,SchroetterI_24a,ZablJ_19a,ZablJ_20a,ZablJ_21a,WendtM_21a,FreundlichJ_21a,LanganI_23a,CherreyM_23a,CherreyM_24a}
and it is organized as follows. In Sect.~\ref{section:megaflow}, we present the survey design and rationale.
For each of the 22 quasar fields, we have obtained  MUSE observations (from 2 to 11hr)  and high-quality UVES spectra, as discussed in section~\ref{section:data}.
In Sect.~\ref{section:catalogs}, we
present the methods used to generate the catalogs of \MgII\ absorbers and galaxies.
In Sect.~\ref{section:properties}, we describe   the physical properties of the galaxies.
In Sect.~\ref{section:results}, we present the main   properties of the host galaxies. In Sect.~\ref{section:discussion}, we compare our sample to others from the literature.
Our conclusions are given in Sect.~\ref{section:conclusions}.
 
 Throughout this paper, we use a $\Lambda$ cold dark matter ($\Lambda$CDM) model with $\Omega_M=0.307$, $\Omega_\Lambda=0.693$ and $H_0=67$~\kms~Mpc$^{-1}$ \citep[`Planck 2015'][]{Planck_15a}.
 At the typical redshift of our survey $z=1$, $1''$ corresponds to 8.23 kpc.
  All magnitudes are in the AB system.

\section{MEGAFLOW survey design and rationale}
\label{section:megaflow}

The low-ionization \MgII$\lla$ doublet seen in quasar spectra has been recognized as a good tracer of the cool
($T\sim10^4$~K)  CGM for close to thirty years \citep[e.g.][]{BergeronJ_91a,BergeronJ_92a,SteidelC_95a,SteidelC_97a,SteidelC_02a}.
However, finding the galaxy counterpart associated with the \MgII\ absorption is often a complicated process.
Before the advent of integral field spectrographs (IFS),  this required deep
pre-imaging in order to identify host-galaxy candidates and expensive spectroscopic follow-up campaigns to secure the host galaxy  identification from its redshift
\citep[e.g.][]{BergeronJ_91a,SteidelC_97a,ChenHW_08a,ChenHW_10a,ChenHW_10b,ChurchillC_13a,NielsenN_13a,NielsenN_13b}.
Still,  the imaging$+$spectroscopy
technique suffers from several disadvantages: 
(i) it is inefficient,   requiring multiple campaigns  for imaging,   redshift identification, and  
kinematics determination \citep[e.g.][]{KacprzakG_11a,HoS_19a};
(ii) it is biased against emission-line galaxies given that 
 pre-imaging necessary for the pre-identification of host-galaxy candidates
is based on the continuum light; and 
 (iii) it is challenging close to the
line-of-sight (LOS) due to the quasar point spread function (PSF).

These shortcomings can be bypassed using Integral Field Unit (IFU) spectroscopy 
such as SINFONI \citep{EisenhauerF_03a} or the MUSE  instrument \citep{BaconR_10a}
 given that an IFU allows the identification of the galaxy counterpart  without pre-imaging as demonstrated in 
\citet{BoucheN_07a}. In addition, IFU observations also provide
the host kinematics and the morphological information which are both easily determined from such 3D data.
Furthermore, 3D data offers the possibility to   easily subtract the quasar's PSF.

The  exquisite sensitivity and large field-of-view ($1'\times 1'$) of MUSE
allows   us  to   detect galaxies further   from the quasar ($30''$ or $\sim$250 kpc 
at $z = 1$). Its broad wavelength coverage   (4700\AA\ to 9300\AA)
 allows us to target quasar
sight-lines with multiple \MgII\ \lla\ absorption lines over the redshift range from 0.4 to 1.5, which are suitable
for the identification of \OII\ emitting galaxies.

The MEGAFLOW survey is aimed at building a statistical sample with at least 100+ galaxy-quasar pairs
 \citep[i.e. 10$\times$ larger than our SINFONI survey,][]{BoucheN_07a}
in order to
allow for a robust analysis of the relation between the absorption and
the host galaxy properties. To this end, we selected
quasars with
(i)  at least three ($N_{\rm abs} \geq 3$) \MgII\lla\ absorption lines at
redshifts between 0.4 and 1.4,
--- ensuring that a sample $\sim 100$ galaxy-quasar pairs could be build with only 20-25 quasar LOS ---
and 
(ii) with a  REW $\Wr\gtrsim0.5$\AA{} from the  \citet{ZhuG_13a} \MgII\
catalog of 100,000 \MgII{} absorption lines extracted from 400,000 quasar  spectra in SDSS (DR12)~\footnote{Available at \url{https://www.guangtunbenzhu.com/jhu-sdss-metal-absorber-catalog}.}.
 The latter criterion ensures that the host galaxies are within 100~kpc from
the quasar LOS (at $z \sim 1$), namely within the MUSE field-of-view, 
given the well known anti-correlation between the impact parameter and rest-frame equivalent width \Wr{} \citep{LanzettaK_90a,SteidelC_95b}.

Ultimately, the MEGAFLOW survey is made of 22 quasar fields listed in Table~\ref{table:qsos}  and the above  selection criteria resulted in
a sample of 79   \MgII\ absorbers with \Wr{}$\gtrsim 0.5$\AA.
This survey is thus optimized to study the properties of galaxies associated with strong \MgII{} absorption lines, but given the wide wavelength range, it has also  allowed us to  perform
galaxy-centered analysis such as the covering fraction analyses of \citet{SchroetterI_21a,CherreyM_23a} and \citet{CherreyM_24a}.  We discuss   the impact of the MEGAFLOW selection criteria on such analyses in section Sect.~\ref{section:fcovering}.

As part of the MEGAFLOW series, in \citet{SchroetterI_16a} and \citet{SchroetterI_19a}, we
presented a preliminary analysis on quasars probing galactic outflows.
In \citet{ZablJ_19a}, we presented a preliminary analysis on quasars probing extended gaseous disks, based on a first version of the galaxy catalog, referred to as `DR1'.
In \citet{ZablJ_20a}, we presented a first tomographic study of an outflow probed by two background sources.
In \citet{ZablJ_21a}, we presented an extended \MgII{} map of the outflowing material around a $z=0.7$ star-forming galaxy.
In  \citet{WendtM_21a}, we presented a first analysis of the metallicity/dust content of the CGM as a function of azimuthal angle.
In \citet{SchroetterI_21a}, we presented a first analysis on the \MgII{} and \CIV{} covering fractions at $z>1$.
In  \citet{FreundlichJ_21a}, we presented a first attempt at detecting the molecular gas content of a subset of galaxies.
In \citet{LanganI_23a}, we presented an analysis of the impact of gas in/outflows on the main-sequence and mass-metallicity scaling relations.
In  \citet{CherreyM_23a}, we presented an analysis of the \MgII{} covering fractions in groups of galaxies.
In \citet{CherreyM_24a}, we presented an analysis of the \MgII{} covering fration of isolated galaxies.

\section{Data}
\label{section:data}

In  Section \ref{section:MUSE}, we describe the MUSE data and data reduction.
In Section \ref{section:uves}, we describe the UVES data reduction.

\subsection{MUSE data}
\label{section:musedata}

\begin{figure}
\centering
\includegraphics[width=6cm]{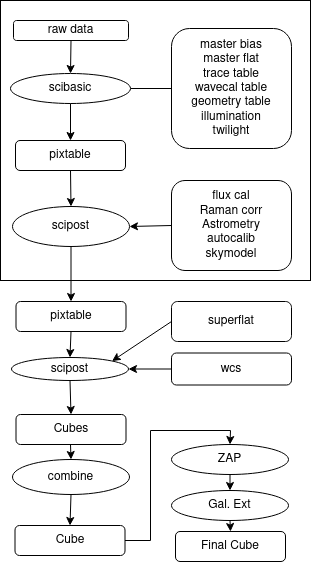}
\caption{Schematic diagram of the steps in the data reduction process. 
}
\label{fig:muse:pipe}
\end{figure}

\label{section:MUSE}

\begin{figure*}
\centering
\includegraphics[width=\textwidth]{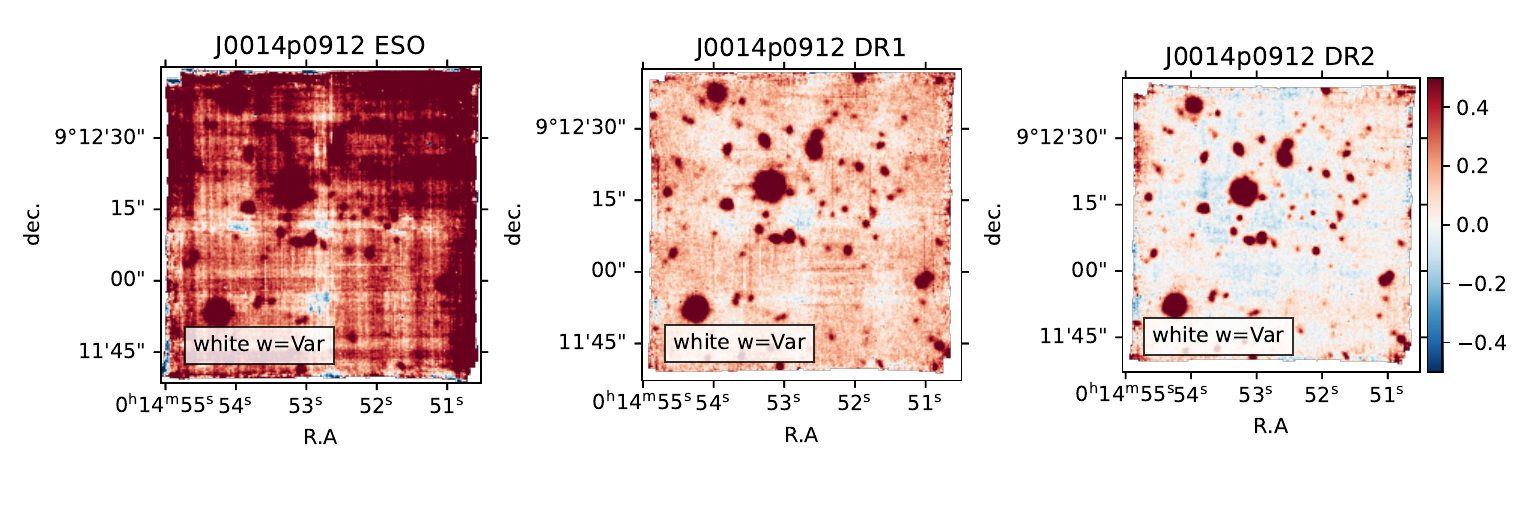}
\caption{White-light images (variance weighted) for the field J0014p0912 from  different data-reductions.
{\it Left}: Automatic data reduction from ESO.
{\it Middle}: Data reduction (`DR1') using custom scripts as in \citet{ZablJ_19a}.
{\it Right}: Final data reduction (`DR2') used here using super flats (Sect.~\ref{section:data}).
}
\label{fig:redux}
\end{figure*}

The MUSE observations were conducted in visitor mode between 2014 and 2018 using guarantee time observations (GTO).
The data were acquired in wide field mode (WFM), using both
standard (WFM-NOAO-N) and adaptive-optics (WFM-AO-N)
modes;  the latter   has been available
since Fall 2017, following the commissioning of the ground
layer adaptive  optics (AO) facility.
Given that the AO observations have  a
gap at 5800–5980 \AA\ due to the AO notch filter, some of the combined (AO$+$ non-AO) datasets have a very heterogeneous PSF in that spectral region.
The program IDs,  total exposure times and the final image quality are listed in \Tab{tab:muse_obs}.

In order to produce a fully calibrated
3D-cube, we used the standard recipes from the MUSE data reduction pipeline 
\citep[version $\geq$2.4,][]{WeilbacherP_14a,WeilbacherP_20a}, and supplemented  some post-processing with custom routines
following \citet{BaconR_17a}, \citet{ZablJ_19a} and \citet{BaconR_23a} with some differences.
The sequence of steps is illustrated schematically in Fig.~\ref{fig:muse:pipe}.


First, raw night calibration exposures were  combined to produce a
master bias, master flat, and trace table (which locates the edges
of the slitlets on the detectors). 
These calibrations were then applied to all the raw science
exposures with the \texttt{scibasic} recipe. Bad pixels corresponding to known CCD defects (columns or pixels) were also masked  
 to reject known detector defects. 
 Subsequently, the \texttt{scibasic} recipe performed the required geometric and wavelength calibrations. At this
point, the pipeline product is a pixel table  (hereafter called \textit{pixtable}) containing all pixel
information: location, wavelength, photon count and an estimate
of the variance.


While the flat-fielding with lamp flats done in the \texttt{scibasic} step removes pixel-to-pixel sensitivity variations, it is not sufficient to ensure an even illumination, especially across the different IFUs. Twilight exposures and night-time internal flat calibrations (called illumination corrections)
are used (when available) for additional  correction in order  to correct for  these flux variations at the slice edges which depends on the ambient temperature. We always used the illumination taken at a similar time and/or with an ambient temperature closer to that of the science exposures.


Next,  the \texttt{scipost} recipe performed the atmospheric dispersion correction, barycentric velocity
correction,   astrometric calibration, telluric correction and flux calibration on the pixtable.
Regarding the flux calibration, observations of spectro-photometric standard stars were reduced in the same way as the science data (save the flux-calibration). The spectral response function used for flux-calibration of the science data was determined by comparing the star's spectrum to tabulated reference fluxes.


As described in \citet[][their Fig.B1]{BaconR_23a},   we can still see a low-level footprint of the instrumental slices and channels
that  arises from  imperfect flat-fielding, which
are difficult to correct for with standard calibration exposures as illustrated on Fig.~\ref{fig:redux}a.
The procedure (referred to as `auto-calibration') to correct for this starts by masking all bright objects in the data using the white-light image
as described in the Appendix B.2 of \citet{BaconR_23a}.
The algorithm calculates the median background flux level in each slice (before sky-subtraction)
using only the unmasked voxels and then scale all slices to the mean flux of all slices
using a robust outliers rejection (15$\sigma$ clipping of the median absolute deviation).
This slice normalization is done in  20 wavelength bins of 200-300\AA\  chosen so that their
edges do not fall on a sky line. This  self-calibration approach is implemented in
the MUSE pipeline for versions $\geq2.4$   \citep[see Sect. 3.10.2 of][]{WeilbacherP_20a}.

Processing the  data  up to this point (box in Fig.~\ref{fig:muse:pipe}) is standard,  performed with the
default parameters of the pipeline within the MuseWise framework~\footnote{\url{http://muse-dbview.target.rug.nl/}.} 
\citep{musewise}.
This stage includes the removal of the sky telluric emission (OH lines and continuum)  and of Raman lines induced by the lasers from the AO facility \citep[][Sect. 3.10.1]{WeilbacherP_20a}.

One remaining imperfection even after the flat fielding steps and the per-slice auto-calibration are sharp flux drops at the edge of those slices that are located at edges of the IFUs.
As discussed in \citet{BaconR_23a} (their Appendix B.4), a way to improve upon these remaining imperfections is through the use of pseudo `skyflats` (or superflats) generated from multiple exposures.
This procedure requires to resample the \textit{pixtable} into a cube sampled  on the same instrumental grid by turning off the dithering and rotations.
This means that in order to create a skyflat for each exposure,  we need to resample all of the others (used as skyflat) to the grid of that exposure.
Given that we need around 30  sky exposures  to create a skyflat,  the required amount of re-sampling is computationally very expensive.

To avoid this  expensive computational effort, we created the pseudo skyflats  from the \texttt{scipost} \textit{pixtable} level, just before resampling the cubes. The pixtables produced by \texttt{scipost} contain fully flux  calibrated - including all flat field steps described above, and we  stacked these pixels. One practical necessity is to mask objects which are in the sky-exposures in the pixtables. The assumption here is that the flat-fielding imperfections, which we aim to catch with the sky-flat, do not shift around on the detector.

The most suited  exposures for constructing the sky-flats are observations of extragalactic deep fields, as they have a relatively sparse source density and, hence, a substantial free sky coverage.
%
%
To have sufficient signal to form a skyflat, we needed 30-50 frames. In total, we constructed  four sky-flats, for the GTO runs in Oct-Nov 2014, Sep-Oct 2015, Jan-Feb 2017, and Aug 2018, from various deep field observations taken as part of MUSE GTO programs, spanning the bulk of the observations.
Once created, we were able to subtract this sky-flat from each of the science exposures in detector-coordinates at the pixtable level.
For   cases where the sky-flats did not perform well, we decided to mask instead the regions with imperfect flat-fielding, again in the pixtable before resampling.
Then, we masked in the pixtable in rare cases individual slices with problems (e.g. due to some problem with the sky subtraction) as well as any satellites that were identified from white-light images create from a first pass reduction.



A datacube was then created
from the skyflat corrected pixtables with the pipeline
recipe \texttt{scipost}, using the default 3D drizzling interpolation process.
We resampled all exposures of a field individually specifying a common output WCS.
We then performed quality control for each exposures, including measurements of the PSF and flux. 
We rejeced any exposures where the flux calibration was off, e.g. due to clouds, or which had a significantly worse seeing than the majority of the exposures in the field.

With the offsets between exposures to correct for the de-rotator wobble, computed on white-light images constructed from the pixtables, we produced data-cubes  resampled to the same WCS pixel grid for each exposures.
These individual data-cubes were finally combined using   MPDAF \citep{mpdaf}.
 This allowed us to perform an inverse-variance weighted average over a large number of data-cubes.
A 3-5$\sigma$ rejection (depending on the number of exposures) of the input
pixels was applied in the average, to remove remaining badpixels and cosmic rays.

The combined datacube was finally processed using the Zurich
Atmospheric Purge \citep[ZAP][]{SotoK_16a} version 2.0~\footnote{\url{https://zap.readthedocs.io/en/latest/}} described in 
\citet[][Appendix B.3]{BaconR_21a}. ZAP performs
a subtraction of remaining sky residuals based on a principal
component analysis (PCA) of the spectra in the background
regions of the datacube.
We provide as input to ZAP an object mask generated from the white-light image of the combined cube  with \textsc{SExtractor} \citep{BertinE_96a}
 to remove the fluxes from all the bright objects.


The cubes were corrected for galactic extinction  
using  the all-sky thermal dust model from the Planck mission \citep{Planck14a}. Specifically, we used the \textsc{dustmaps} Python interface \citep{GreenG_18a}~\footnote{\url{https://dustmaps.readthedocs.io/en/latest/index.html}} to get the galactic color excess $E(B-V)$ from the Planck data. We then corrected the cube using the \citet{CardelliJ_89a} extinction curve with $R_V=3.1$.  This ensures that all spectra and flux measurements in the catalogs will be properly dust corrected.
This cube is referred to as the \texttt{beta} dataset.

In addition, we also produced a version of the cube after subtracting the QSO PSF using
\textsc{Pampelmuse}~\footnote{\url{https://pampelmuse.readthedocs.io/en/latest/}}  \citep{KamannS_13a}
for the non-AO fields and
using the Modelisation of the AO PSF in Python tool
\textsc{MAOPPY}~\footnote{\url{https://github.com/rfetick/maoppy}} 
\citep{FetickR_19a} 
for the fields taken with AO.
This cube is referred to as the \texttt{psfsub} dataset.

\begin{table*}
\centering
\caption{MEGAFLOW quasar fields.
}
\label{table:qsos}
\begin{tabular}{rrrrrrrrr}
\hline \hline
ID~\tablefootmark{a} &
Field name~\tablefootmark{b} & R.A. (J2000)~\tablefootmark{c} & Dec (J2000)~\tablefootmark{d} & $\mathrm{E(B\mbox{--}V)}_\mathrm{gal}$~\tablefootmark{e} & $z_{\rm QSO}$ ~\tablefootmark{f} & $m_r$~\tablefootmark{g} & $N_{\rm abs}$~\tablefootmark{h} & $N_{\rm abs,tot}$~\tablefootmark{i} \\
\hline
11 &
J0014$-$0028 & 00:14:53.36 & -00:28:27.7 & 0.053 &   1.927 & 19.4 & 3 & 7\\
12 &
J0014$+$0912 & 00:14:53.21 & +09:12:17.7 & 0.210 & 2.308& 18.5 & 3  & 13 \\
13 &
J0015$-$0751 & 00:15:35.18 & -07:51:03.1 & 0.038 & 0.875  &  19.3 & 3 & 3 \\
14 &
J0058$+$0111 & 00:58:55.76 & +01:11:28.6 & 0.025 & 1.222 & 18.2  &2  & 3  \\
15 &
J0103$+$1332 & 01:03:32.31 & +13:32:33.6 & 0.036 & 1.663 & 18.6 & 3 & 5\\
16 &
J0131$+$1303 & 01:31:36.45 & +13:03:31.1 & 0.067 & 1.595 & 18.6& 4 & 7 \\
17 &
J0134$+$0051 & 01:34:05.77 & +00:51:09.4 & 0.025 & 1.519 & 18.4 & 4 & 8\\
18 &
J0145$+$1056 & 01:45:13.11 & +10:56:26.7 & 0.068 & 0.938 & 19.1 & 3 & 5 \\
19 &
J0800$+$1849 & 08:00:04.55 & +18:49:35.1 & 0.034 & 1.294 & 17.9 & 4 & 7 \\
20 &
J0838$+$0257 & 08:38:52.05 & +02:57:03.7 & 0.028 & 1.770 & 17.8 & 5 & 8 \\
21 &
J0937$+$0656 & 09:37:49.59 & +06:56:56.3 & 0.044 & 1.814 & 19.3 & 3 & 4\\
22 &
J1039$+$0714 & 10:39:36.67 & +07:14:27.4 & 0.040 & 1.532 & 19.3 & 3 & 5 \\
23 &
J1107$+$1021 & 11:07:42.74 & +10:21:26.3 & 0.026 & 1.925 & 17.6 & 6 & 10\\
24 &
J1107$+$1757 & 11:07:35.26 & +17:57:31.5 & 0.022 & 2.133 & 18.9 & 3 & 8 \\
25 &
J1236$+$0725 & 12:36:24.39 & +07:25:51.5 & 0.022 & 1.605 & 18.7 & 4 & 4\\
26 &
J1314$+$0657 & 13:14:05.62 & +06:57:22.0 & 0.031 &1.880 & 18.0 &  4 & 4 \\
27 &
J1352$+$0614 & 13:52:17.67 & +06:14:33.2 & 0.027 & 1.798 & 18.3 & 3 & 3\\
28 &
J1358$+$1145 & 13:58:09.49 & +11:45:57.6 & 0.021 & 1.484&  18.0 & 4 & 4\\
29 &
J1425$+$1209 & 14:25:38.06 & +12:09:19.2 & 0.027 & 1.618 &  18.5 & 4 & 6\\
30 &
J1509$+$1506 & 15:09:00.12 & +15:06:34.8 & 0.030 & 2.238 &19.3 & 3 & 4\\
31 &
J2137$+$0012 & 21:37:48.44 & +00:12:20.0 & 0.063 & 1.669 & 18.3 & 4 & 5 \\
32 &
J2152$+$0625 & 21:52:00.04 & +06:25:16.4 & 0.060 & 2.389 & 19.4& 4 & 4  \\
\hline
Total & & & &  &  & &  79 &127 \\
\hline
\end{tabular}
\tablefoot{
\tablefoottext{a}{ID of the field}
\tablefoottext{b}{QSO short name.}
\tablefoottext{c}{Right ascension of the QSO [hh:mm:ss; J2000].} 
\tablefoottext{d}{Declination of the QSO [dd:mm:ss; J2000].}
\tablefoottext{e}{Color excess from the Galactic extinction at the position of the QSOs.}
\tablefoottext{f}{QSO redshift $z_{\rm QSO}$.}
\tablefoottext{g}{QSO $r$ magnitude.}
\tablefoottext{h}{Number of \MgII{} absorptions $N_{\rm abs}$ from the SDSS catalog of \citet{ZhuG_13a}.}
\tablefoottext{i}{Total number of \MgII{} absorptions $N_{\rm abs}$ in the UVES data (see Sect.~\ref{section:MgII:catalog}).}
}
\end{table*}

\begin{table*}
\centering
\caption{Summary of MUSE observations for the 22  fields.}
\begin{tabular}{rrrrr}
\hline\hline
Field name~\tablefootmark{a} & $T_\mathrm{exp}$~\tablefootmark{b} & ESO program IDs~\tablefootmark{c} & Instrument Modes~\tablefootmark{d} & PSF~\tablefootmark{e} \\
\hline
J0014$-$0028 &   10.0 & \shortstack{095.A-0365(A), 096.A-0164(A),  0100.A-0089(B),\\ 0101.A-0287(A), 0102.A-0712(A)} & NOAO-N,AO-N & 0.61 \\
J0014$+$0912 &  3.0 & 094.A-0211(B) & NOAO-N  & 0.84 \\
J0015$-$0751 &   3.3 & 096.A-0164(A), 097.A-0138(A), 099.A-0059(A) & NOAO-N & 0.75 \\
J0058$+$0111 &  3.1 &  \shortstack{096.A-0164(A), 097.A-0138(A),\\
0101.A-0287(A), 0102.A-0712(A)} & NOAO-N,AO-N & 0.70\\
J0103$+$1332 &  3.7 & 096.A-0164(A),0100.A-0089(A) & NOAO-N,AO-N & 0.75\\
J0131$+$1303 & 3.7 &  094.A-0211(B),099.A-0059(A) & NOAO-N & 0.81 \\
J0134$+$0051 & 3.6 & 096.A-0164(A), 097.A-0138(A), 099.A-0059(A) & NOAO-N & 0.78\\
J0145$+$1056 &   3.3 & \shortstack{096.A-0164(A), 097.A-0138(A),\\
0100.A-0089(A),0100.A-0089(B)} & NOAO-N,AO-N & 0.69 \\
J0800$+$1849 &  2.0 & 094.A-0211(B) & NOAO-N & 0.56 \\
J0838$+$0257 &   3.3 & 096.A-0164(A),098.A-0216(A) & NOAO-N & 0.54\\
J0937$+$0656 &   11.2 & \shortstack{095.A-0365(A), 0100.A-0089(B),\\ 0100.A-0089(A), 0101.A-0287(A)} & NOAO-N,AO-N & 0.72\\
J1039$+$0714 &   3.3 & 097.A-0138(A),098.A-0216(A) & NOAO-N &0.72 \\
J1107$+$1021 &   3.3 & 096.A-0164(A),098.A-0216(A) & NOAO-N & 0.71 \\
J1107$+$1757 &   2.0 & 095.A-0365(A) & NOAO-N & 0.88 \\
J1236$+$0725 &   5.0 & 096.A-0164(A),0100.A-0089(A),0101.A-0287(A) & NOAO-N,AO-N & 0.73 \\
J1314$+$0657 &   1.7 & 097.A-0138(A) & NOAO-N & 0.53 \\
J1352$+$0614 &   3.6 & 099.A-0059(A),0100.A-0089(A),0101.A-0287(A) & NOAO-N,AO-N & 0.69 \\
J1358$+$1145 &   3.1 & 097.A-0138(A),0100.A-0089(A) & NOAO-N,AO-E  & 0.55\\
J1425$+$1209 &   2.7 & 097.A-0138(A),0100.A-0089(A) & NOAO-N,AO-N & 0.72 \\
J1509$+$1506 &   2.5 & 099.A-0059(A),0100.A-0089(A) & NOAO-N,AO-N & 0.50 \\
J2137$+$0012 &   5.6 & 094.A-0211(B),0102.A-0712(A) & NOAO-N,AO-N & 0.96\\
J2152$+$0625 &  2.0 & 094.A-0211(B) & NOAO-N & 0.57\\
\hline
\end{tabular}
\tablefoot{
\tablefoottext{a}{QSO short name.}
\tablefoottext{b}{Total exposure time  [hr].}
\tablefoottext{c}{ESO program IDs of the MUSE observations.}
\tablefoottext{d}{Instrument mode used (AO/no-AO).}
\tablefoottext{e}{FWHM of the PSF at 7000\AA. }
}
\label{tab:muse_obs}
\end{table*}

\begin{figure}
\centering
\includegraphics[width=9cm]{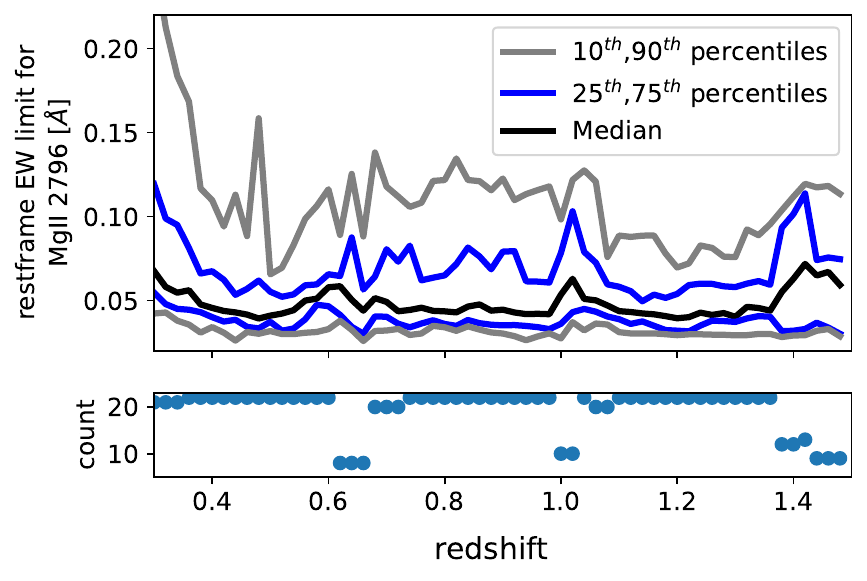}
\caption{Median  3$\sigma$ REW limits for \MgII{} 2796 as a function of redshift for all 22  UVES spectra (solid black line).
The blue and grey lines  represent the 25th and 75th percentiles, respectively.
Figure~\ref{fig:uves:qsoewlimit}  shows the REW limits for each field individually.
 The bottom graph shows the number of sight-lines covering each redshift.
}
\label{fig:uves:meanewlimit}
\end{figure}

\subsection{UVES quasar spectra}

\label{section:uves}

The quasars  of this MUSE GTO-Program were all observed with the high-resolution spectrograph UVES \citep{Dekker:2000a}
between 2014 and 2018 (see Tab. \ref{tab:uves_obs}).
The settings used in our observations were chosen in order to cover the \MgII$\lambda \lambda 2796,2803$ absorption lines as well as other elements such as \MgI$\lambda 2852$, \FeII$\lambda 2586$ when possible. Additional observations were carried out in cycle 108 (2021, 2022) to fill some gaps in redshift coverage for certain sight lines and in particular weak ions. The total observing time amounts to 59 hr (63 hr, including archival observations from 2004).

The data were taken under similar conditions resulting in a spectral resolving power of R $\approx 38000$ dispersed on pixels of $\approx$1.3 \kms. 
The Common Pipeline Language (CPL version 6.3) of the UVES pipeline was used to bias correct and flat field the exposures and then to extract the wavelength and flux calibrated spectra. 
After the standard reduction, the custom software UVES Popler \citep[][version 1.05 from Sept. 2020]{murphy2018popler}  was used to combine the extracted echelle orders into single 1D spectra in vacuum wavelength. The continuum was fitted with low-order polynomial functions \citet{MurphyM_19a}.
 
Figure~\ref{fig:uves:meanewlimit} shows the median 3-$\sigma$ \Wr{} limit as a function of redshift for the 22  quasar spectra (black line).
The gray lines represent the 10th and 90th percentiles, while the blue lines represent the 25th and 75th percentiles.
The \Wr{} limits were obtained from the uncertainty per pixel $\sigma_\mathrm{pix}$, the  average line width of a weak \MgII$\lambda \lambda 2796,2803$ absorption $w_\mathrm{MgII}=5$ \kms, the pixelsize $pix=1.3$ \kms,   local pixel width ${\rm d}l$ in \AA{} and  the detection significance $=3$ using
\begin{equation}
 W_{\rm r}^{2796} \mathrm{limit} = \sigma_\mathrm{pix} \times \sqrt{ \frac{w_\mathrm{MgII}}{\mathrm{pix}}} \times \frac{{\rm d} l}{1+z} \times 3.
\end{equation}

\begin{table*}
\caption{\label{tab:uves_obs} Summary of UVES observations for the MEGAFLOW fields. }
\begin{tabular}{lrrrrrrrrrc}
\hline\hline
QSO field~\tablefootmark{a} & \multicolumn{9}{c}{UVES Grating~\tablefootmark{b}} & ESO program IDs \\
\cline{2-10}\\
  & 346 & 390 & 437 & 520 & 564 & 580 & 600 & 760 & 800 & \\
\hline 
J0014$-$0028 & - & 2.5 & - & - & 2.5 & - & - & - & - & 096.A-0609(A) \\
J0014$+$0912 & - & 0.8 & 1.7 & - & 0.8 & - & - & 1.7 & - & 098.A-0310(A) 096.A-0609(A) \\
J0015$-$0751 & - & 3.3 & - & - & 3.3 & - & - & - & - & 098.A-0310(A) \\
J0058$+$0111 & - & 0.8 & 0.8 & - & 0.8 & - & - & 0.8 & - & 098.A-0310(A) \\
J0103$+$1332 & - & 2.5 & - & - & 2.5 & - & - & - & - & 098.A-0310(A) \\
J0131$+$1303 & - & 1.7 & - & - & - & 1.7 & - & - & - & 096.A-0609(A) \\
J0134$+$0051 & 4.0 & 0.8 & 1.6 & - & 4.0 & 0.8 & - & 1.6 & - & 074.A-0597(A) 098.A-0310(A) \\
J0145$+$1056 & - & 3.3 & - & - & 3.3 & - & - & - & - & 096.A-0609(A) 097.A-0144(A) 098.A-0310(A) \\
J0800$+$1849 & - & 1.6 & - & 1.7 & 1.6 & - & - & - & - & 108.22KC.001 096.A-0609(A) \\
J0838$+$0257 & - & 0.8 & - & - & 0.8 & - & 0.8 & - & - & 098.A-0310(A) 096.A-0609(A) \\
J0937$+$0656 & - & 2.5 & - & - & 2.5 & - & - & - & - & 096.A-0609(A) \\
J1039$+$0714 & 2.5 & 2.5 & - & - & 2.5 & 2.5 & - & - & - & 097.A-0144(A) 108.22KC.001 \\
J1107$+$1021 & - & 3.3 & - & - & 1.6 & 1.7 & - & - & - & 096.A-0609(A) 108.22KC.001 \\
J1107$+$1757 & - & 3.3 & 2.5 & - & 3.3 & - & - & 2.5 & - & 108.22KC.001 096.A-0609(A) \\
J1236$+$0725 & - & - & 1.7 & - & - & - & 0.8 & 1.7 & - & 097.A-0144(A) 096.A-0609(A) \\
J1314$+$0657 & - & 0.8 & - & - & 0.8 & - & - & - & - & 097.A-0144(A) \\
J1352$+$0614 & - & 1.9 & 0.8 & - & 1.9 & - & - & 0.8 & - & 108.22KC.001 097.A-0144(A) \\
J1358$+$1145 & 0.8 & 0.8 & - & - & 0.8 & - & - & - & 0.8 & 097.A-0144(A) \\
J1425$+$1209 & - & 0.8 & - & 0.8 & 0.8 & - & - & - & - & 097.A-0144(A) \\
J1509$+$1506 & - & 2.5 & - & - & 2.5 & - & 1.7 & - & - & 108.22KC.001 097.A-0144(A) \\
J2137$+$0012 & - & 1.7 & - & - & 1.7 & - & - & - & - & 293.A-5038(A) \\
J2152$+$0625 & - & 2.5 & - & 2.5 & - & 2.5 & - & - & - & 293.A-5038(A) \\\hline
\end{tabular}
\tablefoot{
\tablefoottext{a}{Field short name.}
\tablefoottext{b}{Exposure time per grating [hr].}
\tablefoottext{c}{ESO program IDs of the UVES observations.}
}
\end{table*}
\begin{figure}
\centering
\includegraphics[width=9cm]{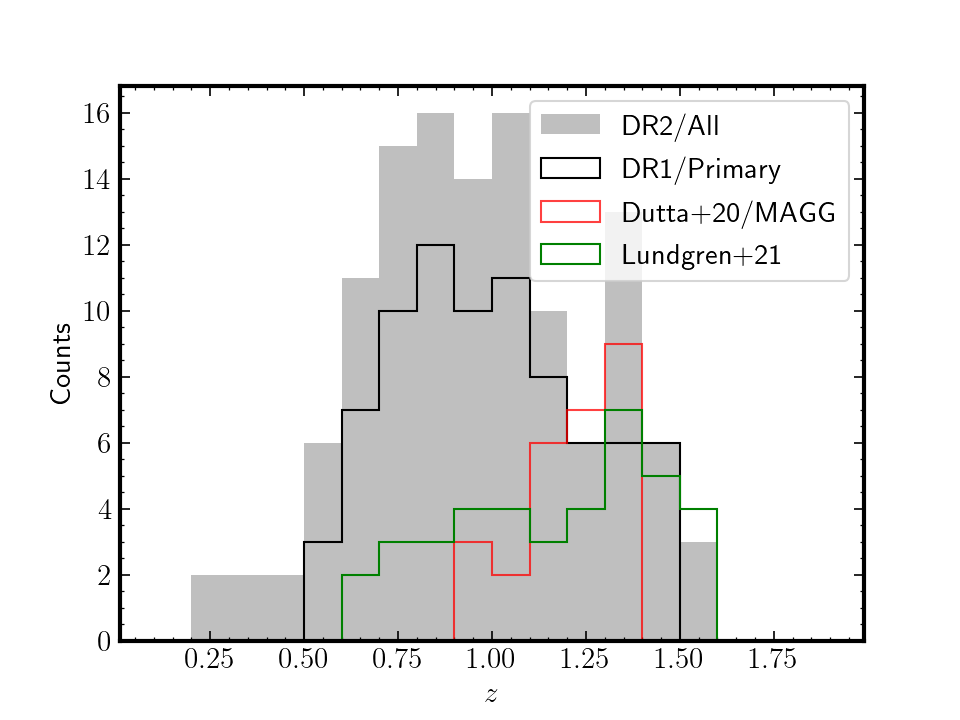}
\includegraphics[width=9cm]{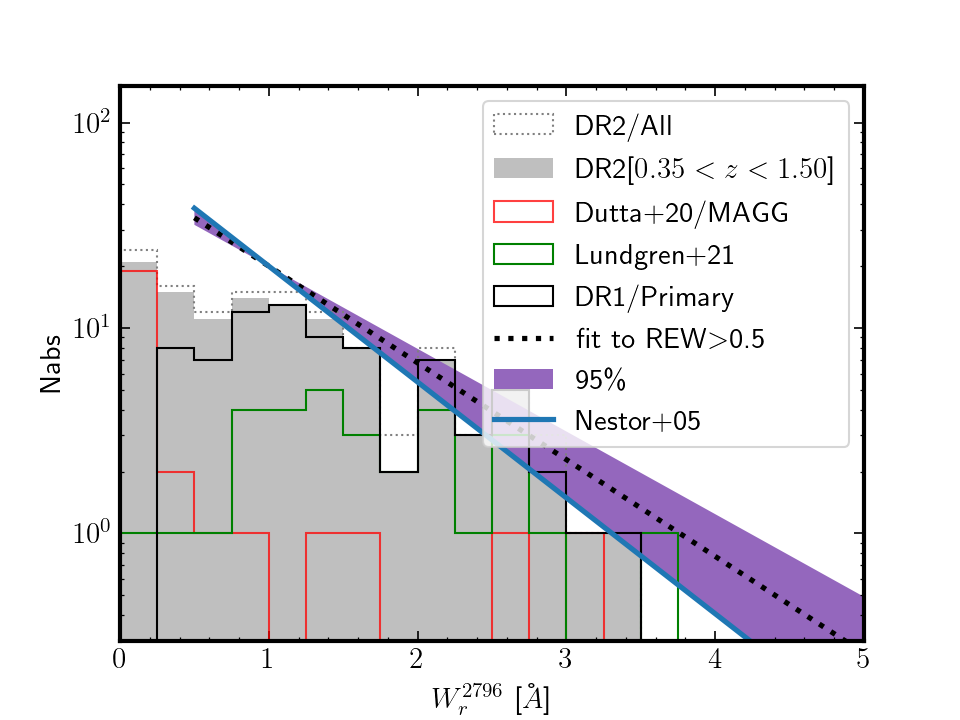}
\caption{Properties of the \MgII\ absorption catalogue. {\it Top}: Redshift distribution of the \MgII{} absorption systems. The solid and gray histograms show the absorbers for the SDSS (DR1) and UVES (DR2)
selections, respectively.
{\it Bottom}: Rest-frame equivalent width \Wr{} distribution of the \MgII{} absorption systems.
The dashed histogram shows the full sample of DR2 absorbers, while the solid and gray histograms represent the absorbers for the SDSS (DR1) and  UVES (DR2) samples, respectively, over the same redshift interval $0.35<z<1.5$.
In order to compare the slope of the \Wr{} distribution to random QSO field, we show the REW distribution from \citet{NestorD_05a} (solid line)
at $z=1$ (normalization arbitrary).
For $\Wr >0.5$~\AA, the 95\%\ confidence interval of the REW slope is shown (tied to $\Wr=1$~\AA). The  \MgII{} REW distribution in MEGAFLOW is not different than from random QSOs for strong \MgII{} absorptions.
In both panels, the red (green) histogram represents the \MgII{} sample from \citet{DuttaR_20a} \citep{LundgrenB_21a}, respectively.
}
\label{fig:uves:rew}
\label{fig:uves:zabs}
\end{figure}

%
%


 
\section{Catalogs}
In this section, we present the
making of the absorption line catalogs
(Sect.~\ref{section:MgII:catalog})
and of the galaxy catalogs (Sect.~\ref{section:catalogs:galaxies}).

\label{section:catalogs}

\subsection{\MgII\ absorption lines}
\label{section:MgII:catalog}

As discussed in Sect.~\ref{section:megaflow}, the SDSS survey selection (DR1) yielded 79 \MgII\ absorption lines in the 22 QSO fields selected from SDSS spectra,  which we  refer to as `DR1' absorber sample and these were used in past studies such as those of \citet{SchroetterI_19a}, \citet{ZablJ_19a} and \citet{WendtM_21a}.
In addition, we searched for serendipitous \MgII\ absorbers in our UVES spectra, and found an additional 48 \MgII{} absorption lines, leading to a total of 127 \MgII{} absorption lines, which is referred  to as the `DR2' sample.

The additional absorption lines were found  independently by the two members of the team (IS and SM). First, the wavelength array of each spectrum is shifted by the rest-frame wavelength ratio (2796.3543/2803.5315)  of the \MgII{}  doublet. The shifted spectrum is then plotted on top of the original spectrum. Such a shift will translate the absorption corresponding to the weaker member of the doublet (\MgII{} 2803) to the location of the stronger member (\MgII{} 2796), since the ratio of observed wavelengths is equal to the ratio of the rest-frame wavelengths, it is independent of   redshift. We took note of such coincidences in each spectrum and determined their redshifts. We then made velocity plots for each of these putative \MgII{} systems to verify the presence or absence of other prominent transitions such as the \FeII,  \CaII, \MnII, and \CIV. However, the presence of these absorption lines was not mandatory for an absorber to be deemed as \MgII. The final \MgII{} catalog was then built by mutual agreement between IS and SM, considering    factors such as the detection significance, velocity structures of the two transitions, and contamination.


The final \MgII{} absorption catalog (DR2) contains 127 \MgII\ absorption lines and the list can be found on the  \texttt{AMUSED}~\footnote{\url{https://amused.univ-lyon1.fr/project/megaflow/}} and MEGAFLOW~\footnote{\url{https://megaflow.univ-lyon1.fr/data}} websites.

 Figure~\ref{fig:uves:zabs} (top) compares the  redshift  distribution of the full \MgII{} absorption line sample (grey histogram) to the pre-selected absorption lines (solid histogram).
Figure~\ref{fig:uves:rew} (bottom) compares the rest-frame equivalent width (REW) $\Wr$ \MgII{} distributions of the full sample (dotted histogram) to the SDSS pre-selected sample (solid histogram). The grey histogram represents the DR2 sample after matching the redshift distribution of the DR1 sample.
For comparison, the \citet{NestorD_05a} and \citet{ZhuG_13a} power laws for the REW distribution of $z=1$ \MgII{} are shown (normalization is arbitrary).
The dotted line represents a fit to the DR2 sample at $0.4<z<1.5$ and for REW greater than $\Wr \geq0.6$~\AA, along with its 95\%\ confidence interval
obtained from 5000 bootstrap fits. The MEGAFLOW \MgII{} absorption lines REW distribution has the same slope as random QSO fields, albeit with a different normalization as pointed in \citet[][their Fig.~A.3]{SchroetterI_21a}. Indeed, the MEGAFLOW $\partial N/\partial W_{\rm r}\propto \exp(-W_{\rm r}/{a})$ with $a=0.86^{+0.15}_{-0.15}$ which is consistent with $a=0.77\pm0.01$ from  field statistics \citep{NestorD_05a,AbbasA_24a} for strong absorbers. 

\subsection{Galaxies}
\label{section:catalogs:galaxies}
In the section below (Sect.~\ref{sect:galaxies:dr1}), we   discuss the
DR1 catalog of \OII{} emitters associated
with the 79 \MgII{} absorption lines.
The DR1 catalog is the basis of the analysis in \citet{SchroetterI_19a}, \citet{ZablJ_19a}, \citet{ZablJ_20a}, 
\citet{ZablJ_21a}, \citet{WendtM_21a} and \citet{FreundlichJ_21a}.
In Sect.~\ref{sect:galaxies:feline}, we discuss our DR2 catalog of all galaxies, which consists of a dual selection of \OII{} line emitters and continuum-selected sources at all redshifts. A preliminary version of this catalog was used in the analysis presented in \citet{LanganI_23a} and \citet{CherreyM_23a}.

\subsubsection{DR1:  {\rm \OII{}} emitters at $z_{\rm abs}$ }
\label{sect:galaxies:dr1}

As discussed in \citet{ZablJ_19a}, the redshift identification of galaxies associated with the 79 \MgII{}
absorption lines was based on multiple pseudo narrow-band (NB) images of width $\approx 400$~\kms{}~\footnote{This filter width gives the optimal S/N for \OII$\lambda\lambda$3727,3729 doublet assuming a line width of FWHM$\approx50$~\kms.} suitable for \OII{}, \OIII{} and \Hb{} emitters
extracted from an early data reduction of the dataset \citep{ZablJ_19a}.
Each NB image is continuum subtracted by using two off-bands, and the NB images are combined in a single NB image (S/N weighted).
We then automatically search for low-S/N objects in these NB images using the detection algorithm \textsc{Sextractor} \citep{BertinE_96a}.
We also searched for quiescent galaxies by creating pseudo-NB images around  the Ca H\&K
doublet.  We ran \textsc{Sextractor} on the inverted NB image given that quiescent galaxies at the right redshift have negative
fluxes in the continuum-subtracted images.
Finally, we  manually checked each candidate. This procedure consisted in a sample of 168 galaxies associated with the 79 absorbers, which  is referred to as the  `DR1' galaxy sample.

\subsubsection{DR2: blind {\rm \OII} emitters}
\label{sect:galaxies:feline}

With $\sim$ 90,000 optical spectra per single pointing, there is demand for
automatic source detections such as the
Line Source Detection and Cataloging (LSDCat) algorithm \citep{HerenzE_17a} or   ORIGIN software \citep{Mary_20a}.
LSDcat is a matched filtering and thresholding algorithm for emission lines algorithm written for IFU data and
used for single emission lines such as \Ly{} emission lines.
After the filtering process, the elements in the data cube represent the statistical S/N an emission line at that voxel position would have
given that it was perfectly represented by the filter template and also assuming fully Gaussian, uncorrelated noise.
The resulting signal is quite robust against varying spectral or spatial sizes of the single-line template.
Hence, LSDcat has also been proven quite effective at indicating other emission features than pure
\Ly. A simple `thresholding-approach', however, is   prone to spurious detections, namely the so-called false positives that directly depend on the threshold.

Here, we use the  {Find Emission LINE objects} (\textsc{FELINE}~\footnote{\url{https://github.com/enthusi/feline}.}) algorithm from \citet{WendtM_24a}  which combines a fully parallelized galaxy line template matching with the matched filter approach for individual emission features of LSDcat.
For the 3D matched filtering, the complete data cube is first median filtered to remove all continuum sources, and then cross-correlated with a template of an isolated
emission feature in two spatial and one spectral dimension. We assumed a simple Gaussian with a FWHM of 250 km/s for the line profile and a PSF based on the given seeing in the data cube. 

The FELINE algorithm then evaluates the likelihood in each
spectrum of the cube for emission lines at the positions provided by a given redshift and a certain combination of typical emission features. FELINE probes all possible combinations of up to 14 transitions paired in 9 groups: \Ha, \Hb, \Hg, \Hd, \OII, \OIII, \NII, \SII, and \NeIII{} for the full redshift range of interest ($0.4 < z < 1.4$). This particular selection of lines is motivated by the most prominent emission features expected in the MUSE data within this redshift range.
This results in 512 ($2^9$) different models that are assessed at roughly 8,000 different
redshifts for each of the $\approx$ 90,000 spectra in a single data cube.
To ensure that only lines above a certain S/N threshold contribute to each
model, a penalty value is subtracted for each additional line. The $S/N$
near strong sky lines are set exactly to that threshold. Therefore, lines that fall
onto such a contaminated region will not affect the model quality. This is
particularly useful for doublet lines that then contribute to a model even
when one of the lines aligns with a skyline.
For each spaxel the model with the highest accumulative probability over all contributing lines and its corresponding redshift are determined.
This approach has the benefit to pick up extremely weak emitters that show
multiple emissions lines while avoiding the deluge of false positives when
looking for single lines below a certain $S/N$ threshold.
This can be applied to each spatial element independently and was thus fully parallelized. From the resulting spatial map of best model probabilities, the peaks were automatically selected via maximum filter and 1D spectra were extracted for each emission line galaxy candidate. Those extracted spectra were fitted with an emission line galaxy template and with the redshift as well as the individual line strengths; the latter  is the only free
parameter to reach sub pixel accuracy in an early redshift estimate and in deriving further diagnostics for the later manual inspection, such as the \OII{} line ratio.

Several of us (JZ, IS, SM, MW, and NB) visually inspected the FELINE solutions with a custom tool and rated each object according to the likelihood of being an \OII{} emitter.  The score was `A' for a clear identification of an \OII{} emitters, `B' for likely \OII{} emitters (due to weak S/N or marginally \OII{} doublet resolved), and `No' for real objects other than \OII{} emitters. A special flag `X' was used when the redshift solution was wrong or needed to be discussed and resolved. The scores (A/B/No/X) were converted to numerical values (2/1/0/-9) and averaged. The FELINE data products are made of the list of FELINE sources (along with the source masks) with their average (and dispersion) \OII\ scores.

\subsubsection{DR2: continuum sources}
\label{sect:galaxies:conti}

We use \textsc{Sextractor} \citep{BertinE_96a} on the white-light images extracted from each datacube to
identify continuum-selected sources. 
For each field, we optimized the detection parameters. Typically, we used a minimum area of 6 to 8 spaxels, a S/N threshold of $\approx1.0$--1.3,
a Gaussian kernel of FWHM 2 spaxels, a deblending threshold of 64 levels, and an automatic background subtraction.

\subsection{Completeness}
\label{section:completeness}

We estimated the completeness of our data by adding $N=150$ fake  \OII\ emitters in two versions of one of our MUSE fields (J0937p0656), using a shallow (2.3~hr) and the deepest (11.2~hr) version.
The \OII\ emitters were placed   at five wavelengths  ($\sim$505, 605, 705, 820 and 920nm) selected to be away from sky emission lines 
and spatially away from known sources.
The emitters are generated using the \gpk{} algorithm using the measured PSF and with realistic galaxy parameters for \OII\ emitters covering a range of redshifts, inclinations, fluxes and surface brightnesses. Specifically, we used an inclination $i$  selected according to a uniform distribution $\sin(i)\propto{\cal U}(0.5,0.95)$ over the range of $i=[30,75]\deg$; position angles uniformly from ${ \cal U}(0,360)$; a  (turbulent) velocity dispersion $\sigma_0=25$~\kms; a  $\tanh$ rotation curve with $V_{\rm max}=100$\kms\ and turn-over radius of $r_v=R_e/1.26$ \citep[following][]{AmoriscoN_10a,BoucheN_15a};  half-light radii $R_e$ of 2.5 and 5~kpc;  \OII\ doublet ratios randomly selected from a normal distribution ${\cal N}(0.8,0.1)$;   flux profiles with S\'ersic index $n=1$; and total fluxes of [1, 2, 4, 6, 8, 10] $\times 10^{-18}$~\flux.  

We ran \textsc{FELINE} on the shallow and deep cubes with fake sources.  The completeness function, $f_{\rm c}$, was then defined as the fraction of sources detected.
Because $f_{\rm c}$ is a complex function of 3D surface brightness in $x,y$ and wavelength,
 it traditionally requires generating a large number of fake sources (over several detection runs), which then requires binning the results in wavelength, sizes, flux, and so on. Instead, we used the Bernoulli regression suitable for binary outcome developed in \citet{BoucheN_19a} and presented in \citet{SchroetterI_21a} which has the advantage that it requires no binning. Briefly, for a set of fake sources, the detectability $Y_i$ is either 0 or 1, and the method consists in parametrizing the completeness function $f_c$, which ranges from 0 to 1, as a Logistic function $L(t)\equiv 1/[1+\exp(-t)]$  where the function $t=F(X_i;\theta)$ is any linear combination of the independent variables $X_i$,  $t=\alpha+\beta_1X_1+\cdots+\beta_n X_n$. The Bayesian algorithm then optimizes the parameters using a Bernoulli likelihood against the observed series of detected and undetected sources $Y_i$. We used the NUTS (No-U-Turn Sampler) from \citet{Hoffman2014}, a self-tuning Hamiltonian Monte Carlo, implemented in the python probabilistic package \textsc{pymc3} \citep{pymc3}. 
 For the logistic function $L$, we use $F$ as a function of flux $f$, size $R$, and wavelength $\lambda$, namely:
\begin{equation}
F(X_i;\theta)
=A[ \log f_i +\hbox{ZP}(\lambda)] + B (\log R_i-\log \overline{R})\label{eq:complete}
\end{equation}
 where $A$ defines the sharpness of the Logistic function $L(t)$, $B$ the size dependence, and  ZP$(\lambda)$ is the wavelength dependent 50\%\ completeness 
taken as a third order polynomial ZP($\lambda)=C+\alpha(\lambda-\lambda_0)+\beta(\lambda-\lambda_0)^2+\gamma(\lambda-\lambda_0)^3$. Here, $\lambda_0\equiv(1+z_0)\times 3727$~\AA\ is the reference wavelength for \OII{} emitters at a redshift of $z_0=1$.

\begin{figure}
\centering
\begin{subfigure}{0.45\textwidth}
\centering
\includegraphics[width=\textwidth]{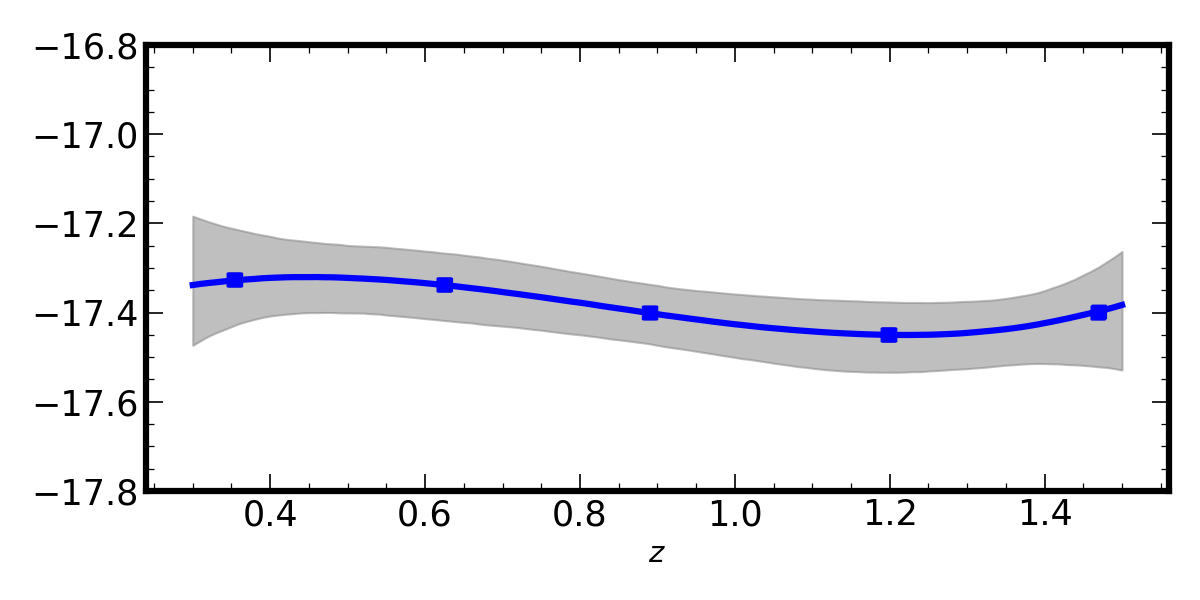}
\caption{}
\end{subfigure}
\begin{subfigure}{0.45\textwidth}
\centering
\includegraphics[width=\textwidth]{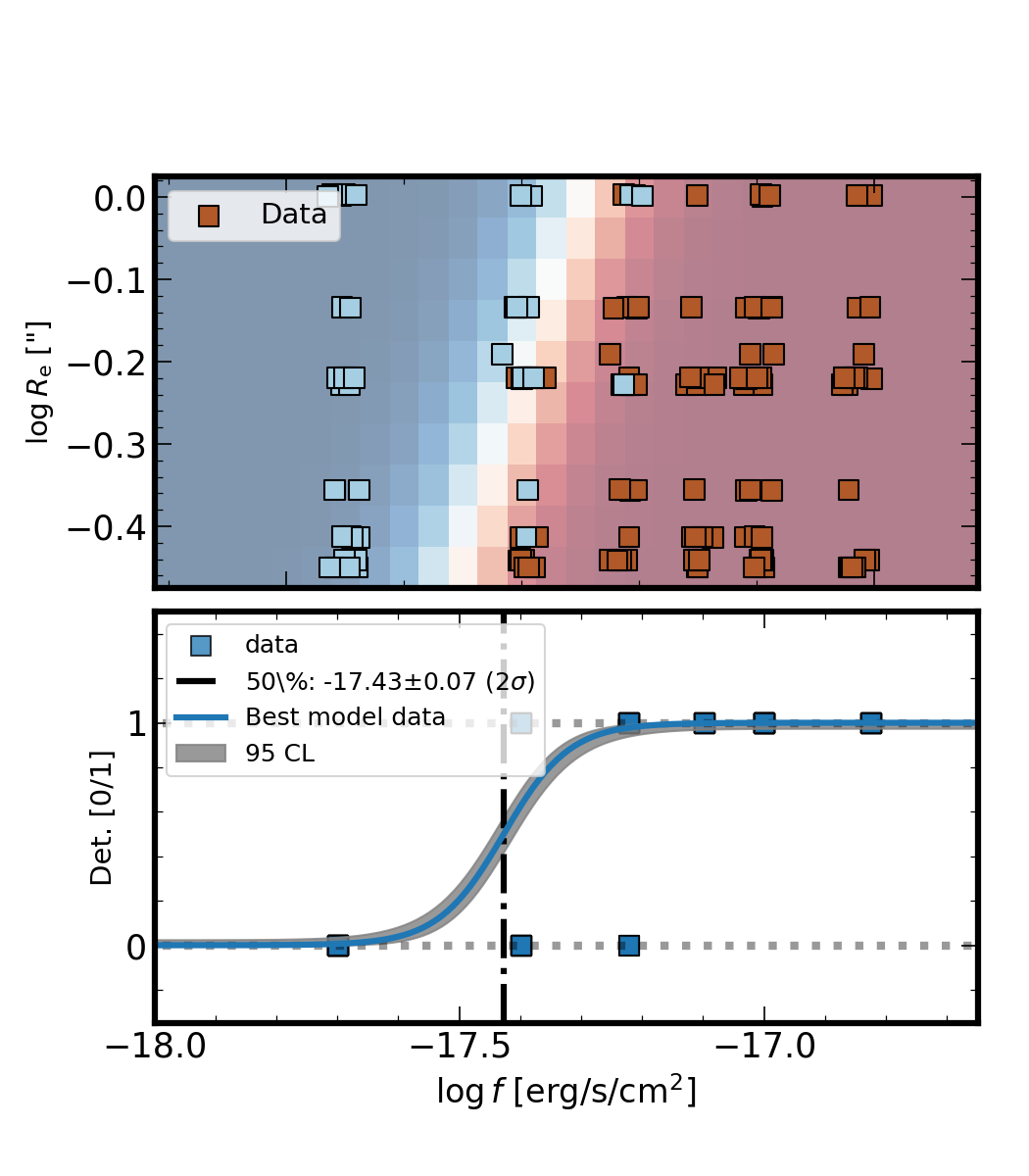}
\caption{}
\end{subfigure} 
\caption{Completeness for \OII{} emitters in the deep cube J0937$+$0656.
(a) Completeness level (50\%) as a function of redshifts.
(b) Completeness as a function of size $R_{\rm e}$ (top panel)
and \OII{} fluxes (bottom panel).
The red (blue) squares represent the \OII{} emitters detected (not detected), respectively.
The shaded area represents the fit to the unbinned data.
}
\label{fig:complete:feline}
\end{figure}

Figure~\ref{fig:complete:feline} shows the completeness achieved for \OII{} emitters in the J0937$+$0656 data as a function of wavelength/redshift (panel a) and as a function of flux and size (panel b). 
The fitted parameters are listed in Table~\ref{table:complete} which indicates that the 50\%{} completeness for \OII{} emitters at $\lambda_0\approx7000$~\AA{} is \detLimOII{} (\detLimOIIshallow) in the deep (shallow) cubes.

\begin{table*}
\centering
\caption{Parameters from the completeness as a function of flux, size, and redshift (see Eq.~\ref{eq:complete}). Errors are 2$\sigma$ (95\%).}
\label{table:complete}
\begin{tabular}{ccccccc}
cube  & A & B  & C & $\alpha$ & $\beta$ & $ \gamma$ \\
\hline
2.3hr & $11.69^{+4.45}_{-3.48}$ &$-9.66^{+5.25}_{-5.86}$ & $-17.15^{+0.09}_{-0.09}$ & $-0.05^{+0.30}_{-0.32}$ & $-0.16^{+0.54}_{-0.54}$ & $-0.17^{+1.20}_{-1.18}$\\
11hr & $18.45^{+5.84}_{-5.94} $ & $-8.57^{+6.68}_{-7.45}$   & $ -17.43^{+0.07}_{-0.08} $ &$-0.20^{+0.26}_{-0.27}$ &$0.29^{+0.46}_{-0.46}$ & $0.58^{+1.02}_{-1.05}$ \\
\hline
\end{tabular}
\end{table*}

\begin{figure*}
\centering
\includegraphics[width=9cm]{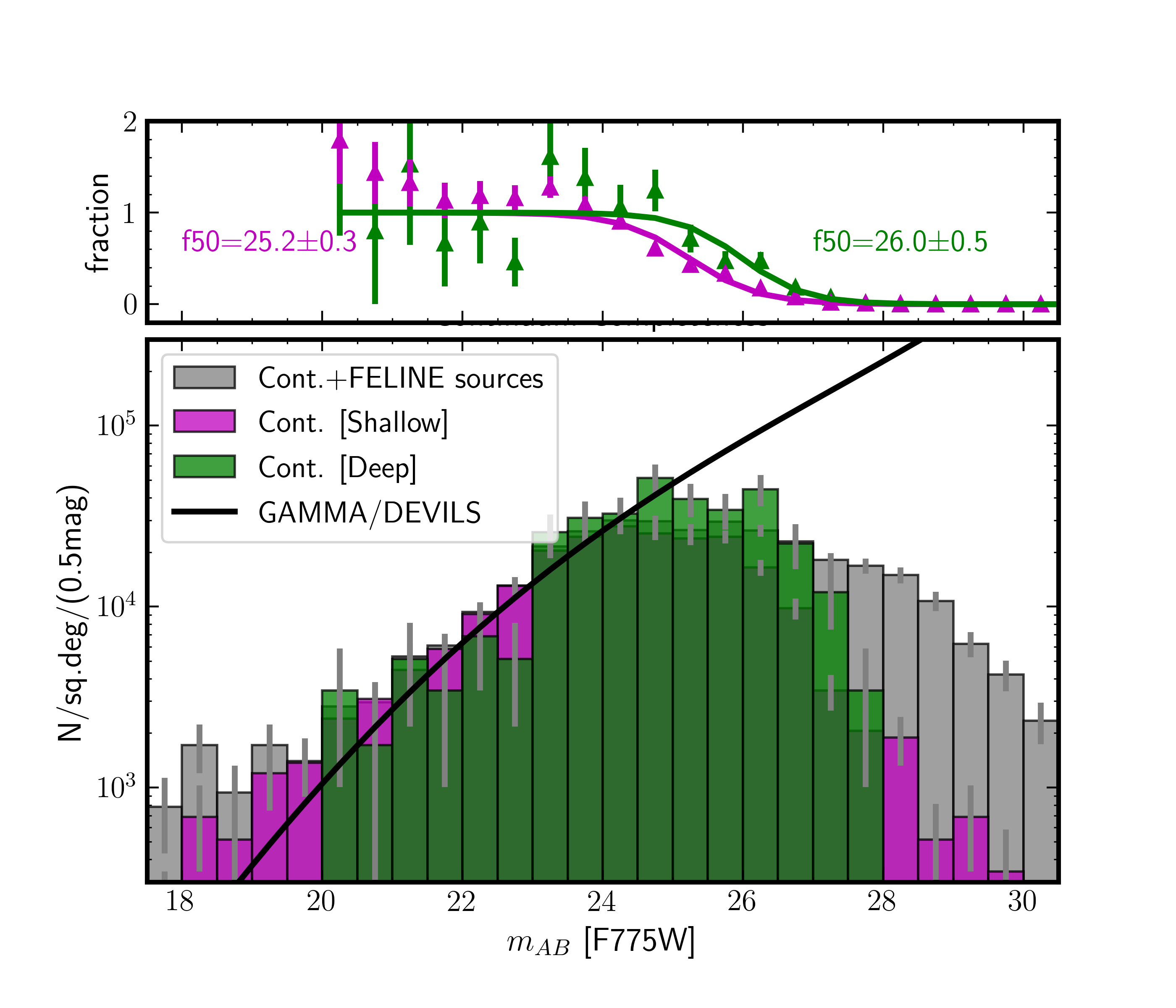}
\includegraphics[width=9cm]{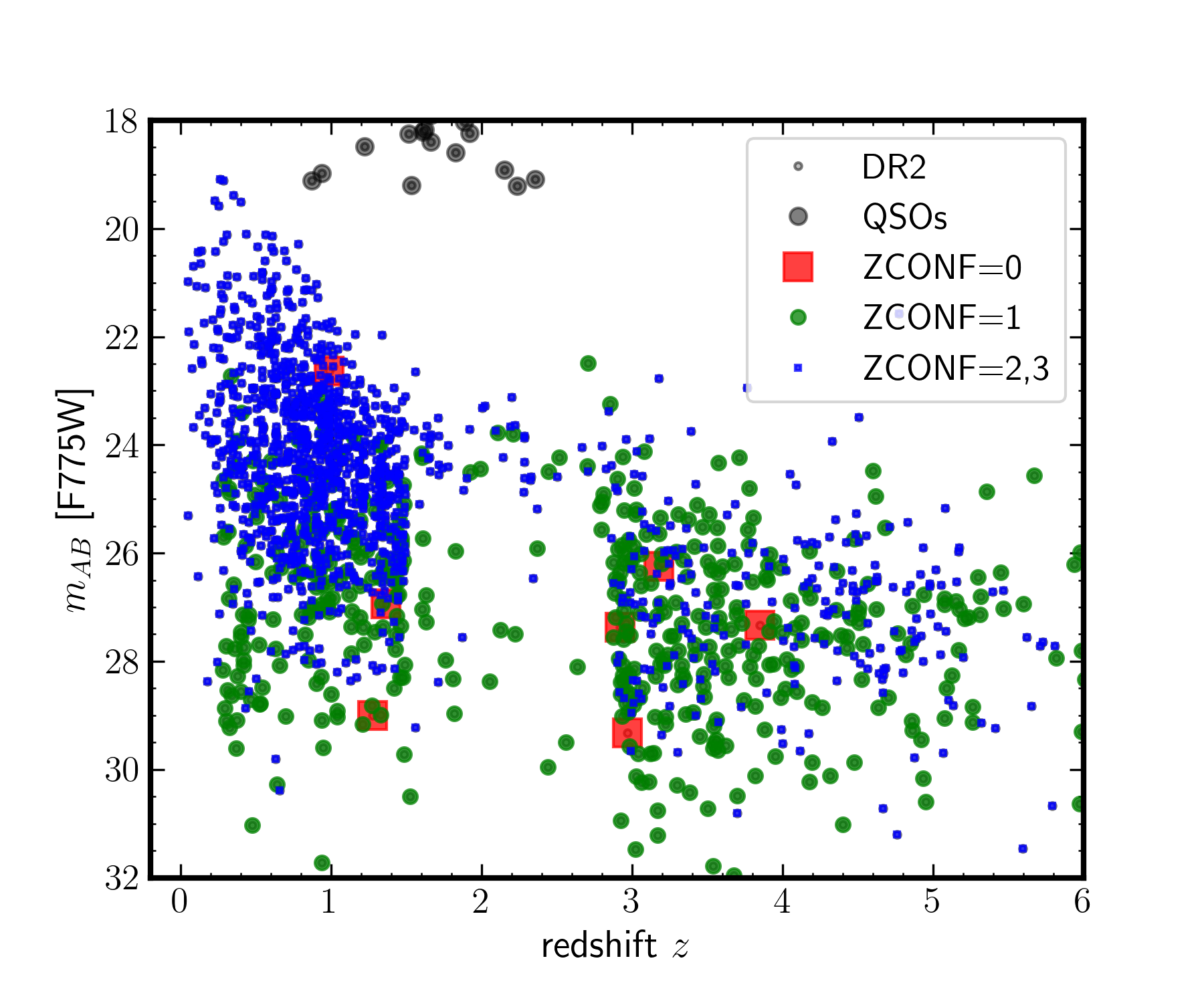}
\caption{Magnitude distributions for the DR2 catalog. {\it Left:} Magnitude F775W number counts. 
The grey histogram represents the entire MEGAFLOW survey. The magenta histogram represents the 20 shallow fields and the green histogram the 2 deep fields.
	Compared to the number counts from the GAMMA/DEVILS survey (solid line) from \citet{devils}, the MEGAFLOW survey is 50\%-complete to $i\approx \detLimContshallow$ (\detLimCont)  in the shallow (deep) fields, respectively. 
{\it Right:}  Magnitude-redshift distribution of all sources with redshifts $z>0$. The QSOs are shown as grey circles. The galaxies with \texttt{ZCONF}$=1$ (2,3) are represented as green circles (blue squares), respectively. 
}
\label{fig:complete:white}
\end{figure*}

Regarding the completeness of continuum sources, we estimated the completeness from the number counts 
($N/{\rm deg^2}/{\rm 0.5 mag}$) shown in Fig.~\ref{fig:complete:white}(left). 
By fitting the F775W number counts normalized to the expected counts from large galaxy surveys such as the recent GAMMA/DEVILS survey \citep{devils}, we find that
the 50\%\ completeness is  $m_{F775W}\approx\detLimContshallow$ the shallow fields (2-4hr)
and is $m_{F775W}\approx \detLimCont$  for the two deep fields (11hr).


\subsection{Source inspection}

At this point each source in the FELINE and continuum-based catalogs are assigned a 6-digit \verb|FELINE_ID| or \verb|WHITE_ID| using the following nomenclature:
`xxa001'. Here, `xx' is the unique field ID (Table~\ref{tab:muse_obs}) and `a' is 0 for FELINE source, 1 for continuum sources,  and 2 for continuum sources near the QSO, in the \verb|psfsub| dataset.

Following \citet{BaconR_23a}, we used the \textsc{pyMarZ} software \citep{Hinton_16a}
 to identify five redshift solutions for the FELINE and continuum sources with 9 templates for passive and star-forming galaxies
and then used \textsc{pyPlatefit}~\footnote{A python-inspired version of the \textsc{Platefit} IDL code  \citep{BrinchmannJ_04a}.}
 to perform emission and absorption line fitting on each of the redshift solution.
We also computed the narrow-band images associated with each emission or absorption lines with $S/N>2$
packaged in `source' fits files,  which are important to assess the detection of weak lines.
The FELINE, \textsc{pyMarZ} redshift solutions, the fitted lines with $S/N>2$, the narrow-band images and spectra are all used to produce static html files as in \citet[][their Fig.19]{BaconR_23a}.

These static html files are then fed into a modified version of the \textsc{SourceInspector} tool \citep{BaconR_23a}, 
a Python-Qt interface that allow users to select one of the redshift solutions or to provide a new one. However, most importantly, the \textsc{SourceInspector} tool allows the user to match 
continuum-detected sources with FELINE-detected sources (and vice-versa) with confidence.

\subsection{Categories for inspection}
\label{sec:inspec:categ}

At this stage, the catalog contains a high fraction of false detections to make sure it is highly  complete. The removal of the false detections requires an inspection.
In order to optimize the time required for inspecting the catalogs  with \textsc{SourceInspector}, we pre-matched the FELINE and continuum sources and focused on galaxies at $0.4<z<1.4$ that are likely to either be star-forming galaxies with \OII{} or passive galaxies with just continuum.
Specifically, we cross-matched the FELINE and continuum catalogs according to their RA, DEC and defined the following categories:
\begin{itemize}
\item 1.1 `GoodFelineCont'  sources with FELINE score $\geq1.0$ and continuum detection;
\item 1.2 `GoodFelineNoCont'  sources with FELINE score $\geq1.0$ and no continuum detection;
\item 1.3 `GoodFelineNoMarz' sources with FELINE score$\geq1.0$ whose redshift is not among the 5 \textsc{Marz} solutions within 150~km/s;
\item 2.1 `BrightContGoodFeline' bright continuum sources with magAUTO$<24.5$ (25.0 for the  two deep fields);
\item 2.2 `BrightContNoFeline' bright continuum sources without a FELINE match;
\item 3.0 `FelineBorderLine' sources with FELINE score between 0.5 and 1.5; or with a score standard dispersion $>0.45$
\item 4.0 `Other'  sources not in the other categories
\end{itemize}

For instance, for the deep field J0014$-$0028, there are 289 FELINE sources and 251 continuum sources, and they are distributed in each category  as follows:
\begin{itemize}
\item 1.1 GoodFelineCont 66
\item 1.2 GoodFelineNoCont 23
\item 1.3 GoodFelineOffMarz 18
\item 2.1 BrightContGoodFeline 30
\item 2.2 BrightContNoFeline 31
\item 3.0 FelineBorderline 46
\item 4 Other (FELINE) 180; and Other (Cont) 190
\end{itemize}

Several of us (MW, IS, JZ, RB, NB, SM, and JR) inspected the redshift solutions for the categories 1.x, 2.x and 3.0 with \textsc{SourceInspector} where each member assessed the redshift confidence (\textsc{ZCONF}) ranging from 0 to 3.


A confidence level of \texttt{ZCONF=0} corresponds to the situation where no redshift solution was found.
A confidence level of \texttt{ZCONF=1} corresponds to a low confidence solution, namely, when the redshift solution remains uncertain (could be \OII\ or \Ly) owing to the low S/N of the line.
A confidence level of \texttt{ZCONF=2} corresponds to a high S/N single line whose shape is identifiable (e.g. a resolved \OII{} doublet, an asymmetric \Ly).
A confidence level of \texttt{ZCONF=3} corresponds to a secure redshift with multiple lines.

 \begin{table}
 \centering
 \caption{Distribution of redshift confidence \texttt{ZCONF} for various classes of sources. \label{table:zconf}}
 \begin{tabular}{lccccc}
 \hline
 & \multicolumn{4}{c}{\texttt{ZCONF}} \\
 Class  & 0&  1 & 2 & 3 & All \\
 \hline
 QSOs & 0 & 0  &0 & 22 & 22\\
 Stars ($z=0$) & 0 & 0 & 3 & 54 & 57 \\
 low$-z$ ($0<z<0.35$) & 0 & 17 & 2 & 90 & 109 \\
 \OII ($0.35<z<1.5$) & 6 & 173 & 244 & 734 & 1157 \\
 Desert ($1.5<z<2.8$) & 0 & 28 & 20 & 28 & 76 \\
 LAE ($z>2.8$) & 7 & 376 & 276 & 10 & 669 \\
 unknown & 337 & 0 & 0 & 0  & 337 \\
 All & 350 & 594 & 545 & 938 & 2427 \\
 \hline
 \end{tabular}
 \end{table}

\subsection{Reconciliation}

The results of the visual inspections were then combined and when there were several different redshift solutions proposed, we resolved the disagreement during the reconciliation meetings.

\begin{figure}
\centering
\includegraphics[width=6cm]{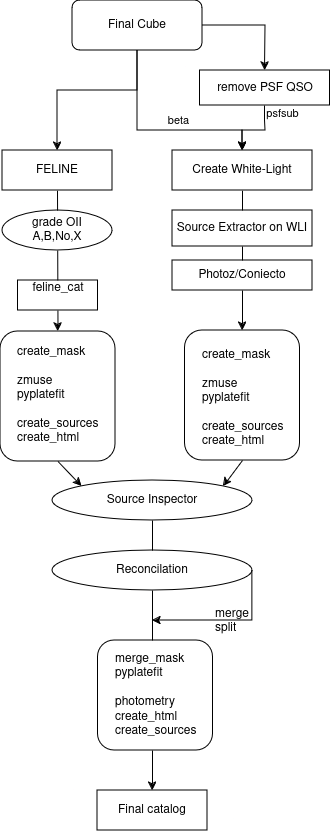}
\caption{Schematic illustration of the process used to generate the final catalog from the FELINE and continuum-based catalogs.}
\label{fig:inspections}
\end{figure}

The inspection and reconciliation process also yielded some sources that required to be split (in cases where the \textsc{SExtractor} deblending failed) or to be merged (in cases where FELINE assigned two redshifts due to a large kinematic gradient). 
 The entire sequence of steps in the process were processed with custom routines forming the {MegaFlow Catalog Processor (\textsc{MFCP})} which is based on 
 \textsc{MuseX} used in \citet{BaconR_23a}.
 The sequence is illustrated in Fig.~\ref{fig:inspections}.

 As discussed in Sect.~\ref{sec:inspec:categ}, the entire process is optimized for galaxies at $z<1.5$  and is somewhat biased against \Ly{} emitters that fall into the category `4', which were not systematically inspected.

\subsection{Final DR2 catalog}

The  final catalog  (v2.0) contains 2427 sources, which includes 22 quasars, 57 stars,  1998 galaxies with \texttt{ZCONF}$\geq1$, and 350 sources with no redshifts (\texttt{ZCONF}$=0$).
The statistics of \texttt{ZCONF} in the final catalog  is given in Table~\ref{table:zconf} for various classes such as nearby galaxies, \OII{} emitters, galaxies in the redshift desert ($1.5<z<2.8$), and \lya{} emitters at $z>2.8$.  

Prior to performing the photometric measurements,  the object masks  were  merged into a single mask for the sources with both FELINE and continuum detections. 
We note that this leads to a series of object masks which can be overlapping.  For instance, the QSO source was obtained from the \verb|beta| dataset, while nearby sources were obtained from the \verb|psfsub| dataset; in this situation, the small sources can be embedded into the QSO segmentation mask.
The are also emission line objects that are in the foreground/background from a continuum object. In this situation, the object masks can overlap significantly.
As a result, we provide a \verb|is_blended| flag for all sources.

Using \textsc{Photutils} (v1.4.0) and theses object masks, we determined  the photometry for each objects in $R$, $I$, $SDSS_r$, $SDSS_i$, F775W, and
13 pseudo-medium bands (320\AA{} wide) covering the wavelength range from 4780 to 9260 \AA, excluding the laser notch filter.
The  magnitude--redshift distribution for the 1998 galaxies in the DR2 final catalog with redshift $z>0$  is shown in Fig.~\ref{fig:complete:white} (right). The graph clearly shows the redshift gap at $1.5<z<2.8$ due to the lack of lines between \lya{} and \OII{} in the MUSE wavelength range.
The magnitude completeness limits as a function of redshift confidence 
\texttt{ZCONF} are given in Table~\ref{tab:redshift:complete} 
using the method described in Sect.~\ref{section:completeness}.

In  Sect.~\ref{section:compareDR1}, we compare the properties of the DR2  galaxy catalogs to the previous DR1 version.

\begin{table}
\centering
\caption{Magnitude (F775W) completeness limit (\mAB) for the deep (shallow) fields, respectively.
\label{tab:redshift:complete}
}
\begin{tabular}{lccc}
\hline
 & 50\% & 90\% \\
\hline
All &  26.0 (25.2) & 25.0 (24.2)\\
\texttt{ZCONF}$>=1$ &  25.7 (25.0) & 24.8 (23.9) \\
\texttt{ZCONF}$>=2$ & 25.5 (24.8) & 24.7 (23.8) \\
\hline
\end{tabular}
\end{table}

\subsection{Final DR2 data products}

A final five-digit ID was assigned to each source with the convention that `xx001' was reserved for the QSO, where `xx' is the field ID from Table~\ref{tab:muse_obs}. 
The catalog files are described in Appendix~\ref{appendix:catalogs}.

We also provide  
  new htmls and new source fits files   for each object, which are available on the \texttt{AMUSED}  interface~\footnote{\url{https://amused.univ-lyon1.fr/project/megaflow/}} after running \textsc{PyPlatefit} in
order to perform emission and absorption line fitting on each emission or absorption line independently. 
An example of a final html file is shown in  Fig.~\ref{fig:examples}.
 The \texttt{AMUSED} interface is searchable and details are presented in \citet{BaconR_23a}.

\section{Physical properties}
\label{section:properties}

\subsection{SED stellar masses}
\label{section:SED}

With the photometry and the 13 pseudo-medium bands,
 we applied the custom SED fitting code \textsc{Coniecto} as in \citet{ZablJ_19a} to all sources with $m_r<26.5$.
This code is described in \citet{ZablJ_16a}. 
In brief, we used the BC03 models \citep{BruzualG_03a}  with a delayed-$\tau$ star-formation history (SFH) and nebular line $+$
continuum emission added following the recipe by  
\citet{SchaererD_09a} and \citet{OnoY_12a}. Here, we used a \citet{ChabrierG_03a} 
 unitial mass function (IMF) and a \citet{CalzettiD_00a} extinction law.
 While we
used the same extinction law both for nebular and stellar emission,
we assumed higher nebular extinction, $E_N (B - V)$, than stellar
extinction, $E_S (B - V)$, with  $[E_S (B - V ) = 0.7 E_N (B - V )]$.

\begin{figure*}
\centering
\includegraphics[width=7.5cm]{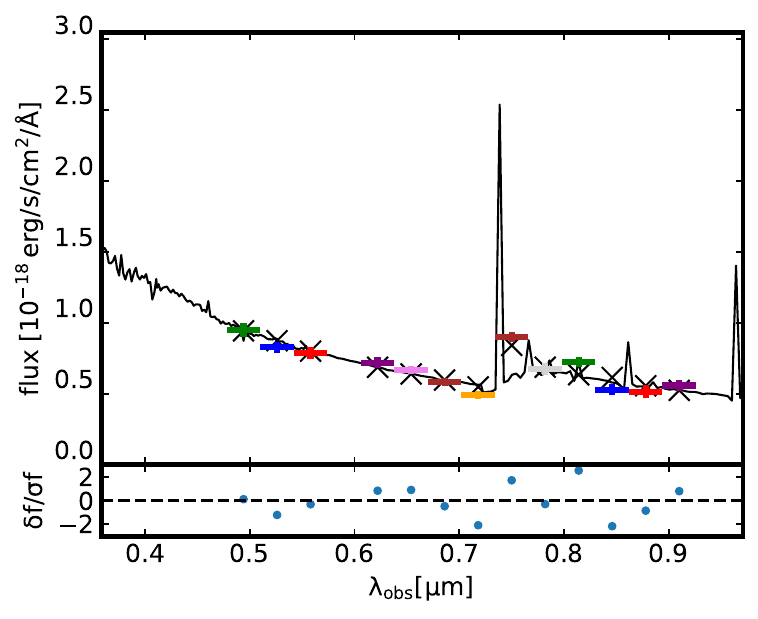}
\includegraphics[width=8.5cm]{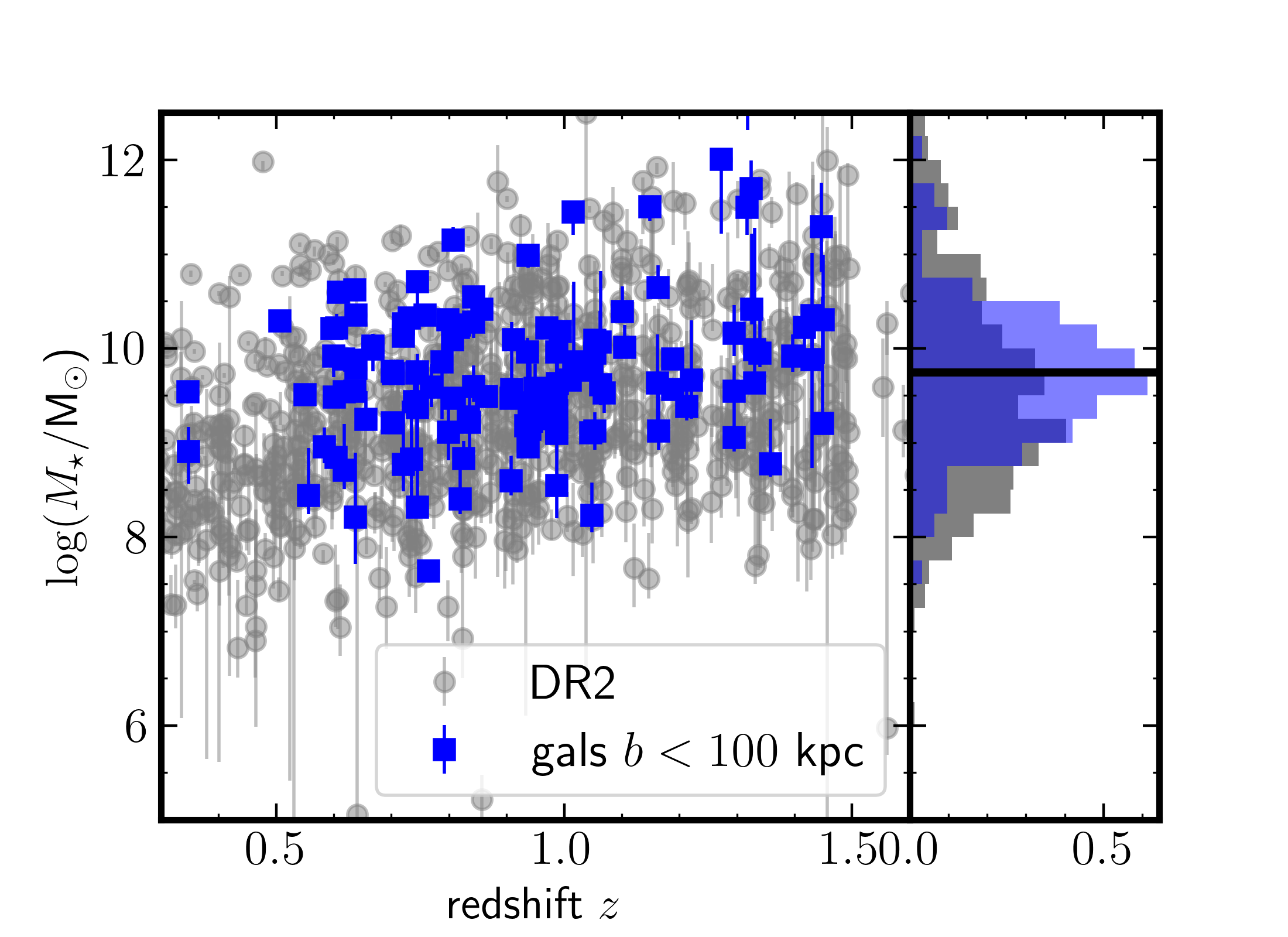} 
\caption{Stellar mass estimations. {\it Left:} Spectral Energy Distribution (SED) for galaxy ID=22055 at $z=0.9856$ from field J1039p0714. 
The black line represents the best fit template. The solid bars represent the pseudo-narrow bands.
{\it Right:}
Stellar-mass vs redshift for the galaxies in the DR2 catalog at $0.35<z<1.5$.
The  squares represent the galaxies within 500~\kms\ of the \MgII{} absorbers and with impact parameters $b<100$~\kpc. The histogram shows the normalized stellar mass distributions.  
}
\label{fig:sed:mass}
\end{figure*}

Fig.~\ref{fig:sed:mass} shows an example of SED fit for galaxy ID=22055 with a redshift of $z=0.9856$, 
and stellar mass of $M_\star=10^{9.18}$~\msun\ and
 the stellar mass-redshift distribution   for the low-redshift galaxies in the DR2 catalog (right panel).
The filled squares in Fig.~\ref{fig:sed:mass}(right) show the  mass-redshift distribution for the galaxies within 100~kpc of the QSO sight-lines. The histogram inset of Fig.~\ref{fig:sed:mass} (right) shows that the \MgII{} host galaxies have a mean (median) stellar mass of $\log M_\star/\msun=$ 9.75 (9.67), respectively.

\subsection{Star formation rates}

We compute dust-corrected SFRs using the \OII{} fluxes using the procedure presented in \citet{LanganI_23a}, which is based on the
empirical calibration of \citet{GilbankD_10a,GilbankD_11a}   that incorporates the dependence of the Balmer decrement with stellar mass.
Specifically, we used
\begin{equation}
\hbox{SFR}=\frac{\hbox{L}_{\OII}/(1.12\times2.53\times10^{40} \hbox{erg~s$^{-1}$})}{a\times\tanh[(x-b)/c]+d}
\end{equation}
where \LOII{} is the \OII{} luminosity, $x=\log M_\star$,  $a,b,c=-1.424,9.827,0.572$ and $d=1.70$. The factor 1.12 converts the
\citet{GilbankD_10a} calibration to the \citet{ChabrierG_03a} IMF used throughout the survey.

\subsection{Morpho-kinematics}

In order to  derive the morpho-kinematic of the galaxies, we used the \gpk algorithm \citep{BoucheN_15a} which performs a forward fit of a disk model with 10 free parameters directly on the 3D MUSE data. \gpk takes into account the effect of the spectral line spread function (LSF) and of the point spread function (PSF) to deconvolve the observations and yields the intrinsic galaxy properties. These include the main axis orientation,   inclination,   half-light radius,   maximum velocity,   velocity dispersion, along with the  flux and   position of the galaxy. We ran \gpk on subcubes centered on either \OII, \OIII{} or \Ha\ emission lines, with the continuum subtracted, using the PSF from Sect.~\ref{section:musedata} and the LSF as in \citet{BaconR_23a}; namely, the median LSF FWHM is LSF[\AA]$(\lambda)=5.866\times10^{-8}\lambda^2 - 9.187\times10^{-4}\lambda+6.040$ where $\lambda$ is in \AA.


\section{Results}
\label{section:results}
\subsection{Matching galaxies with absorbers}
\label{subsection:primaries}

\begin{figure*}
\centering
\includegraphics[width=8cm]{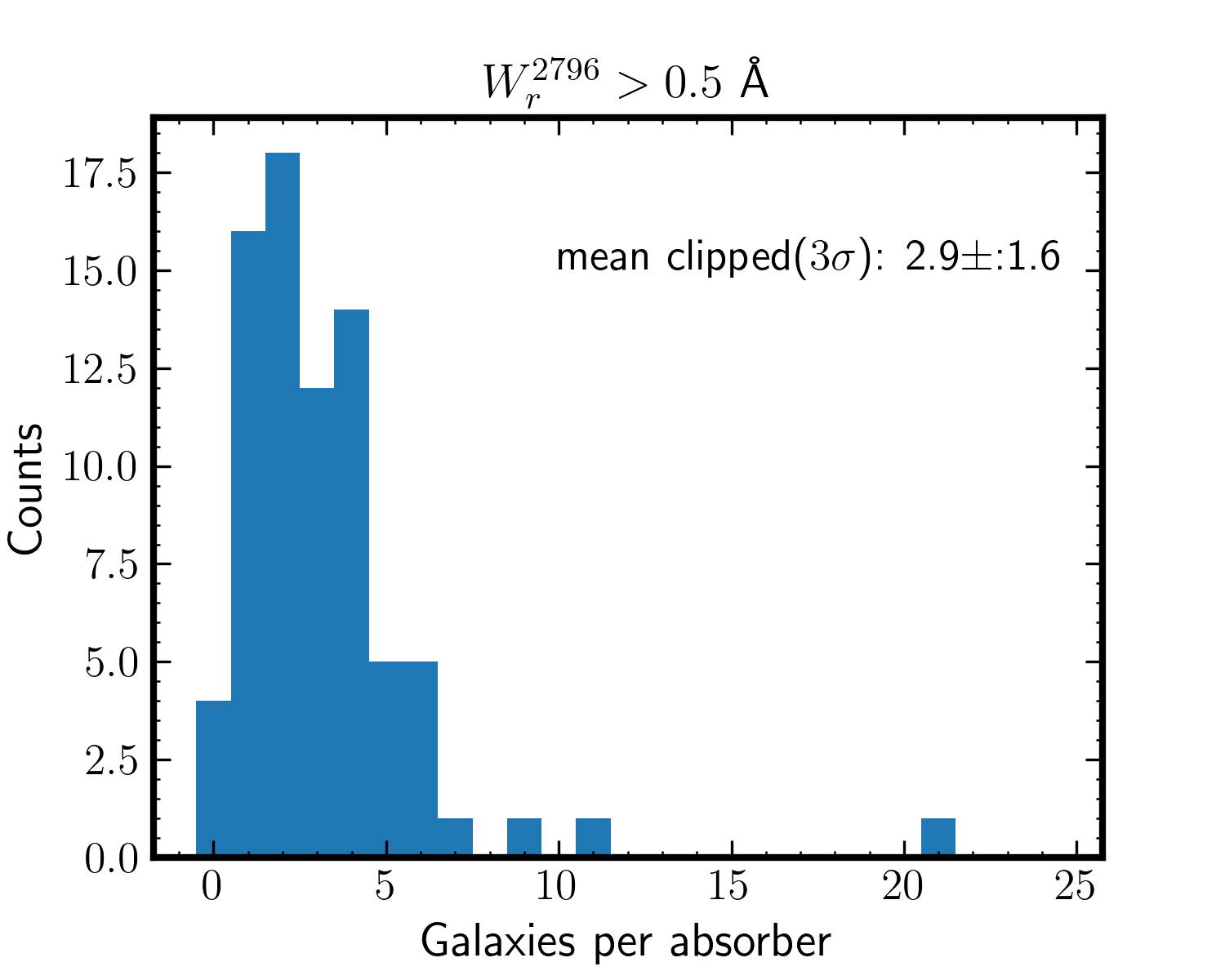}
\includegraphics[width=8cm]{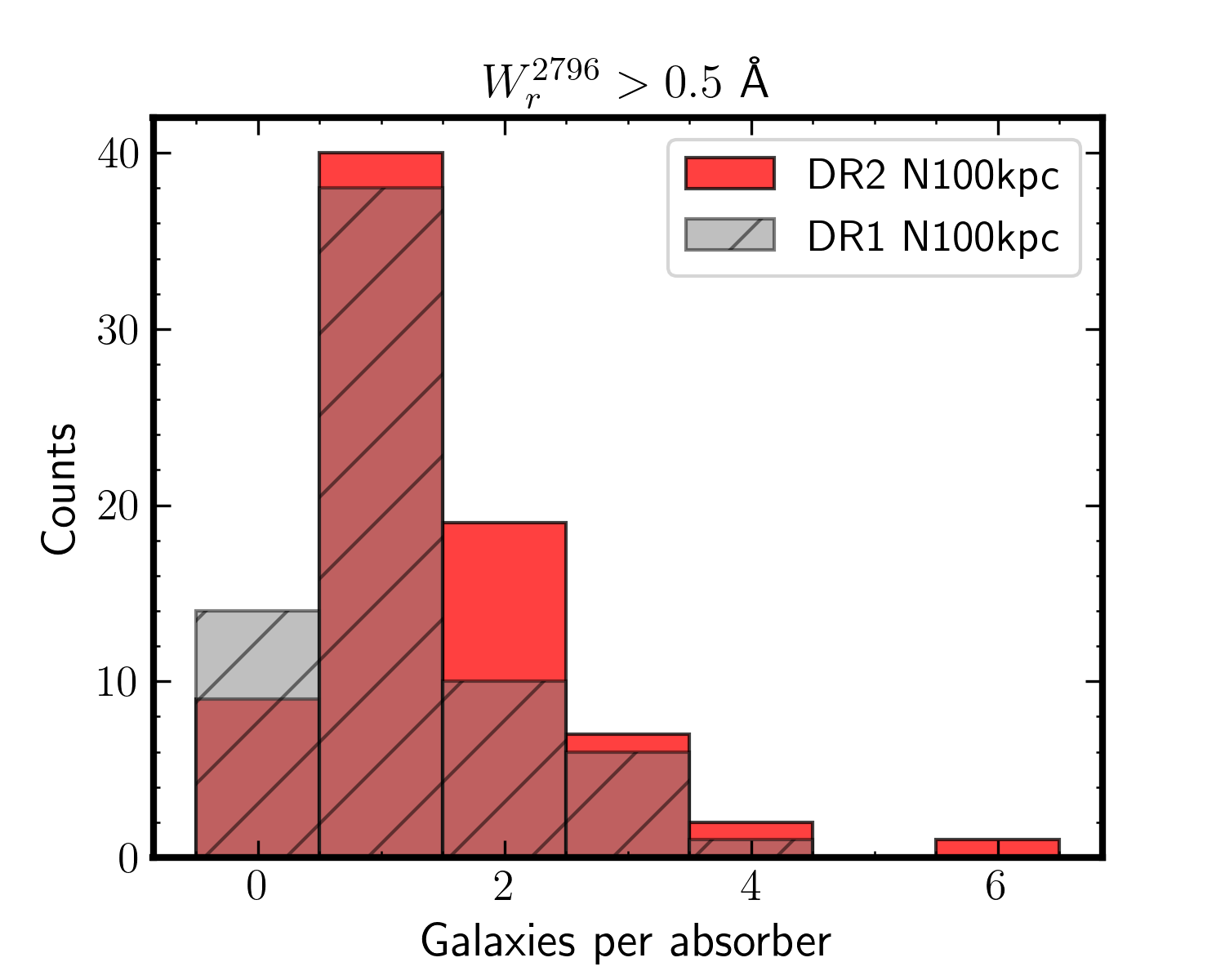}
\caption{{\it Left:} Distribution of the number of galaxies  that are within $\pm$500~\kms\ of strong ($\Wr>0.5$~\AA) absorption systems, within the MUSE FOV, and
within the MUSE wavelength coverage for \OII{} emitters, i.e. $0.35<z_{\rm abs}<1.45$.
The mean (median) number of galaxies is 2.9$\pm$1.6 (3.0), respectively.
{\it Right:} The red (hashed) histogram shows the number of galaxies   around strong ($\Wr >0.5$~\AA) \MgII{} absorption systems within 100~kpc in the redshift range $0.35<z<1.45$ suitable for \OII{} emitters, from the DR2 (DR1) samples, respectively. 
}
\label{fig:N_per_abs}
\end{figure*}

The association of absorption systems with their galaxy counterparts is crucial to understanding the physical processes at play in the CGM.
We  first associated the galaxies to \MgII{} absorption lines according to the redshift difference between the galaxy and the absorption, $\Delta v$.
Using  a $\Delta v=500$~\kms,   the number of galaxies (in the entire MUSE FOV) per \MgII{} absorption system was derived.
From its distribution  shown in  Fig.~\ref{fig:N_per_abs}(left), we can see that there is approximately an average of
\begin{equation}
N_{\rm obs}=2.9\pm1.6
\end{equation}
 galaxies per absorber. As discussed in \citet{CherreyM_23a}, 
this is close to the expected mean number of galaxies, namely,
\begin{equation}
N_{\rm exp}= 3.3\pm3.1,
\end{equation}
 in a cylinder of radius of $r_{\rm max}<280$~kpc, (corresponding to the same area of the MUSE FOV) and height  of $\Delta z=500$~\kms\ for galaxies in halos of mass $M_{\rm h}>10^{11}$\msun{} using the halo mass function from \citet{TinkerJ_08a}.
 We note that  the drop in Fig.~\ref{fig:N_per_abs}(left)  at  $N=5$ is entirely caused by the MUSE FOV.

If we restrict ourself to galaxies within an impact parameter $b$ less than 100~kpc,   the red-filled histogram in 
Fig.~\ref{fig:N_per_abs}(right)  shows the number of galaxies  that are within $b<100$~kpc  around 80 strong ($\Wr >0.5$~\AA) absorption systems. This figure shows that the majority (60 out of 80, i.e. 75\%) of \MgII{} systems are matched with one or two galaxies, and only ten have no counterparts.

\begin{figure}
\centering
\includegraphics[width=0.9\columnwidth]{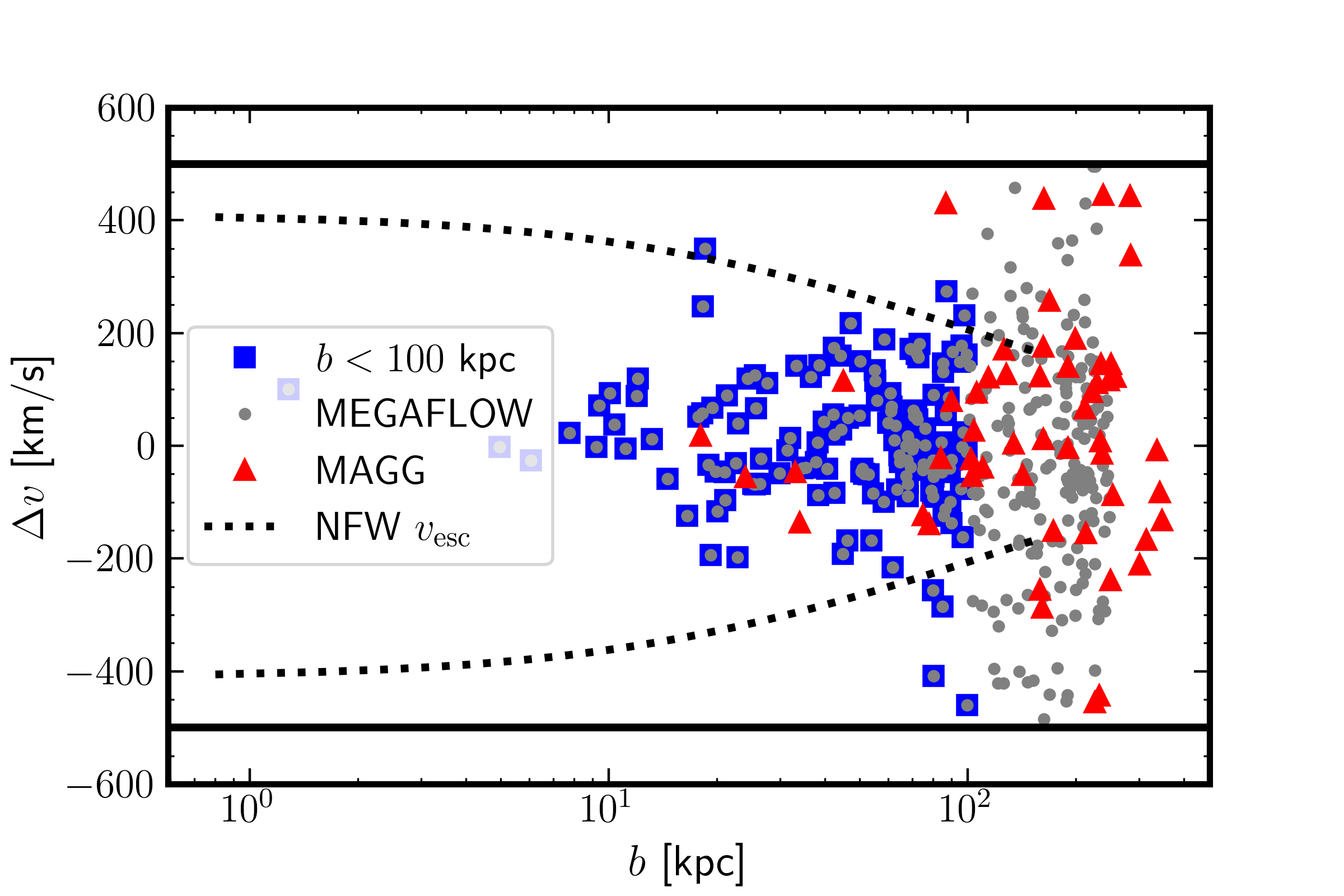}
\caption{Velocity offset $\Delta v$ versus impact parameter $b$.
The squares show the MEGAFLOW sample for galaxies within the impact parameter of $b< 100$ or $b<$250 kpc.
The  triangles represent the MAGG sample from \citet{DuttaR_20a}.
The dotted line represents the escape velocity $v_{\rm esc}$ for the median halo mass of isolated galaxies $M_h=10^{11.7}$~\msun\ \citep{CherreyM_24a}
}
\label{fig:dvdr}
\end{figure}

Figure~\ref{fig:dvdr} shows the velocity difference ($\Delta v$) between the galaxy redshift and the absorption redshift as a function of  the impact parameter, $b$.
The $\delta v$  standard deviation is $\sigma\sim100$~\kms, much smaller than our search boundary of $\pm 500$~\kms, as in \citet{HuangY_21a}.
This shows that, in spite of our search range of $|\Delta v|<500$~\kms{} (solid horizontal lines), the galaxies associated with the \MgII{} absorbers are found within $\pm 200$~\kms{}
from the absorption lines and within $\pm 100$~\kms{} at distances less than 50~kpc. 
The dotted line represents the escape velocity $v_{\rm esc}$ of a $10^{12}$ \citep[][NFW]{NavarroJ_97a} halo, and
this figure indicates that the \MgII{} gas is not virialized as  $\Delta v$ decreases towards the center, whereas it would increase  as in groups \citep{CherreyM_23a} if the gas were virialized. It shows also that we are not likely affected by mis-assigned gas, namely within $|\Delta v|<500$~\kms{} but outside the virial radius. This was discussed in \citet{HoS_20a} in the context of the \MgII{} in EAGLE simulations or in \citet{WengS_24a} in the context of Lyman limit systems in TNG50.

\subsection{Comparison between DR1 and DR2 galaxy catalogs}
\label{section:compareDR1}

A notable difference between our DR1 and DR2 galaxy catalog is that 
the DR1 galaxy catalog from \citet{ZablJ_19a} was constructed primarily from pseudo-narrow band images at the redshift of \OII{} corresponding to the \MgII{} absorption catalog,
while the final DR2 catalog is  constructed in such a way that it is totally blind to the presence of \MgII{} absorptions.

Out of the 80 (69) systems in DR2 (DR1), there are 115 (80) galaxies within 100 kpc associated with absorbers in the DR2 (DR1) samples, respectively.
This corresponds to a   `success' rate in finding at least one galaxy of $\sim90$\%{} (see Table~\ref{tab:success}) for strong \MgII{} systems ($\Wr >0.5~$~\AA), compared to DR1 which had $\approx80$\%\ success rate \citep{SchroetterI_19a,ZablJ_19a}.

Fig.~\ref{fig:N_per_abs}(right) compares the number of galaxies per absorber within $b<100$~kpc, for strong ($\Wr >0.5$~\AA) \MgII{} systems.
The red (grey) histogram represents the number of galaxies per absorption lines in the DR2 (DR1) catalogs.
This figure also shows that there are fewer systems (10 vs 14) without galaxies, while there are more systems  (80 vs 58) with one or two galaxies.

\begin{table}
\centering
\caption{Success rate in finding the host galaxy with \OII{} within the MUSE wavelength coverage, i.e. at $0.35<z_{\rm abs}<1.5$. 
}
\label{tab:success}
\begin{tabular}{lrrrrr}
\hline\hline
$\Wr$ (\AA) \tablefootmark{a} & $N_{\rm abs}$ \tablefootmark{b} & $N_{100}\geq1$ \tablefootmark{c} &  $f$(\%) \tablefootmark{d} & $N_{\rm gals}^{100}$ \tablefootmark{e}   \\
\hline
$>$0.2 & 95 & 74 & 0.78 & 119    \\
$>$0.5 & 78 & 69 & 0.88 & 113  \\
$>$0.8 & 64 & 56 & 0.88 & 89   \\
$<$0.2 &   20   & 10   &     0.5      &  17 \\
\hline
 All &126 & 85  & 0.67 &138  \\
\hline
\end{tabular}
\tablefoot{
\tablefoottext{a}{Minimum \MgII{} REW.}
\tablefoottext{b}{Number of absorptions lines.}
\tablefoottext{c}{Number of absorptions lines with at least 1 galaxy with 100~kpc ($N_{100}\geq1$.}
\tablefoottext{d}{Fraction of \MgII{} absorptions with $N_{100}\geq1$.}
\tablefoottext{e}{Number of galaxies within 100 kpc}
}
\end{table}

\begin{figure*}
\centering
\includegraphics[width=18cm]{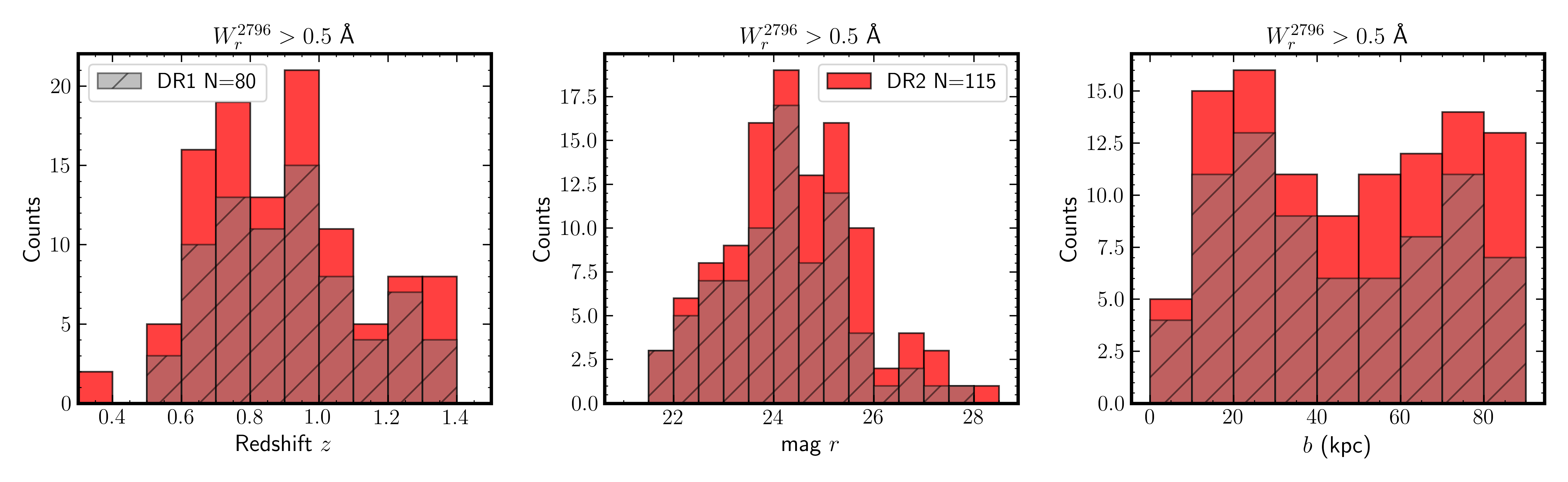}
\caption{Comparison between the DR2 and DR1 galaxy sample of galaxies within $\pm$500~\kms\ around each \MgII{} absorption system and within 100~kpc.
The left, middle and right panels show the redshift, magnitude and impact parameter distributions for galaxies in the redshift range $0.3<z_{\rm abs}<1.5$, respectively.
}
\label{fig:compareDR2DR1}
\end{figure*}

Figure~\ref{fig:compareDR2DR1} compares the properties of the DR2 and DR1 catalogs, in particular it compares the redshift, magnitude and impact parameters of the galaxies associated with \MgII{} absorptions at $0.3<z<1.5$ in the DR1 and DR2 catalogs.
The difference between the red solid and hashed histogram shows the benefit from a blind approach.

\subsection{Primary galaxy}

{Within the MEGAFLOW sample, we define {'primary'} galaxies as the ones we could unambiguously identify  as responsible for \MgII\ absorption in quasar spectra (if any at the same redshift). We identify these primary galaxies blindly (without considering \MgII{} absorptions) by applying the following criteria:}

\begin{itemize}
    \item $0.3 < z < 1.5$;
    \item ZCONF $\geq$ 2;
    \item z$_{\rm gal}$ $<$ z$_{\rm QSO}$ with $\Delta v_{\mathrm{QSO}}\geq1000$ \kms;
    \item  $b < 150 $ kpc;
    \item smallest $b$ within $\pm 1000$~\kms. Any neighbor must have $b$ at least 50 kpc greater;
    \item N$_{\rm FOV}$ < 5 to avoid groups.
\end{itemize}
 
Here, $z_{\rm gal}$ and $z_{\rm QSO}$ are, respectively, the redshift of the galaxy and the redshift of the quasar of the field, $\Delta v_{\mathrm{QSO}}$ as the velocity difference between the quasar and the galaxy and N$_{\rm FOV}$ being the number of galaxies in the MUSE field of view in a $\pm 1000\kms$ window.
{In total, 170 \textit{primary} galaxies have been identified. Finally, out of the 115 absorptions in the redshift range $0.3<z<1.5$, 43 can unambiguously be associated with a  {primary} galaxy. On the other hand 127  {primary} galaxies are not associated with any absorption.}

This set of  {primary}  galaxies (associated or not associated with an absorption) are particularly useful in improving our understanding of the physical parameters responsible for the presence of  an absorption at a given impact parameter.

\subsection{SFR distribution}


We investigated the impact of the SFR on the presence of a counterpart \MgII\ absorption by comparing the SFR distribution for  {primary} galaxies (as described in \ref{subsection:primaries}) associated with an absorption to those not associated with an absorption. Figure \ref{fig:MS_abs_vs_noabs} shows the stellar mass versus SFR, namely the main sequence for these two sub-samples.
We can see that  this MUSE survey is sensitive to galaxies with $M_\star\gtrsim10^{7.5}$ and SFR $\gtrsim0.01$~\mpy{}, but that the majority of \MgII{} host galaxies have SFR$>1$~\mpy{} and $M_\star>10^9$~\msun.
This indicates that the survey would have detected  satellites with mass ratio $1:20$ or greater.

\begin{figure}
 \centering
\end{figure}

\begin{figure*}
\centering
 \includegraphics[width=8.cm]{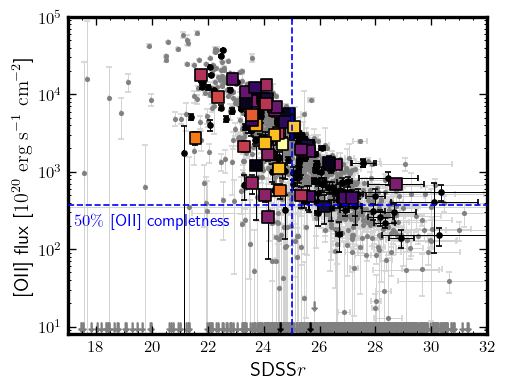}
\includegraphics[width=8.5cm]{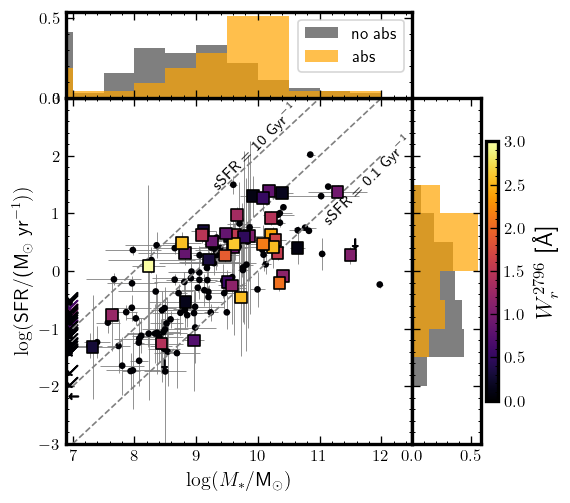}
\caption{Star formation rates. {\it Left:} Measured \OII{} flux versus SDSS $r$ magnitude recomputed with \textsc{photutils}. The full MEGAFLOW sample is represented with gray dots.
Colored and black dots represent respectively  {primary} galaxies that are associated and not associated with a counterpart absorption. Galaxies with no detected \OII{} flux are represented by downward arrows. {\it Right:} Estimated stellar mass versus estimated SFR for the  {primary} galaxies. The top and right histograms present respectively the stellar mass distribution and the SFR distribution for the galaxies associated (orange) and not associated (gray) with a \MgII\ absorption. Arrows on the left indicates galaxies without stellar mass estimation and/or no SFR estimation (because no \OII\ emission detected). The primary galaxies associated with an absorption are colored according to the \MgII{} absorption rest-frame equivalent width. Error bars are 1$\sigma$ uncertainties.
}
\label{fig:OII_vs_SDSSr}
\label{fig:MS_abs_vs_noabs}
\end{figure*}



\subsection{Gas density profile}

\begin{figure}
\includegraphics[width=9.5cm]{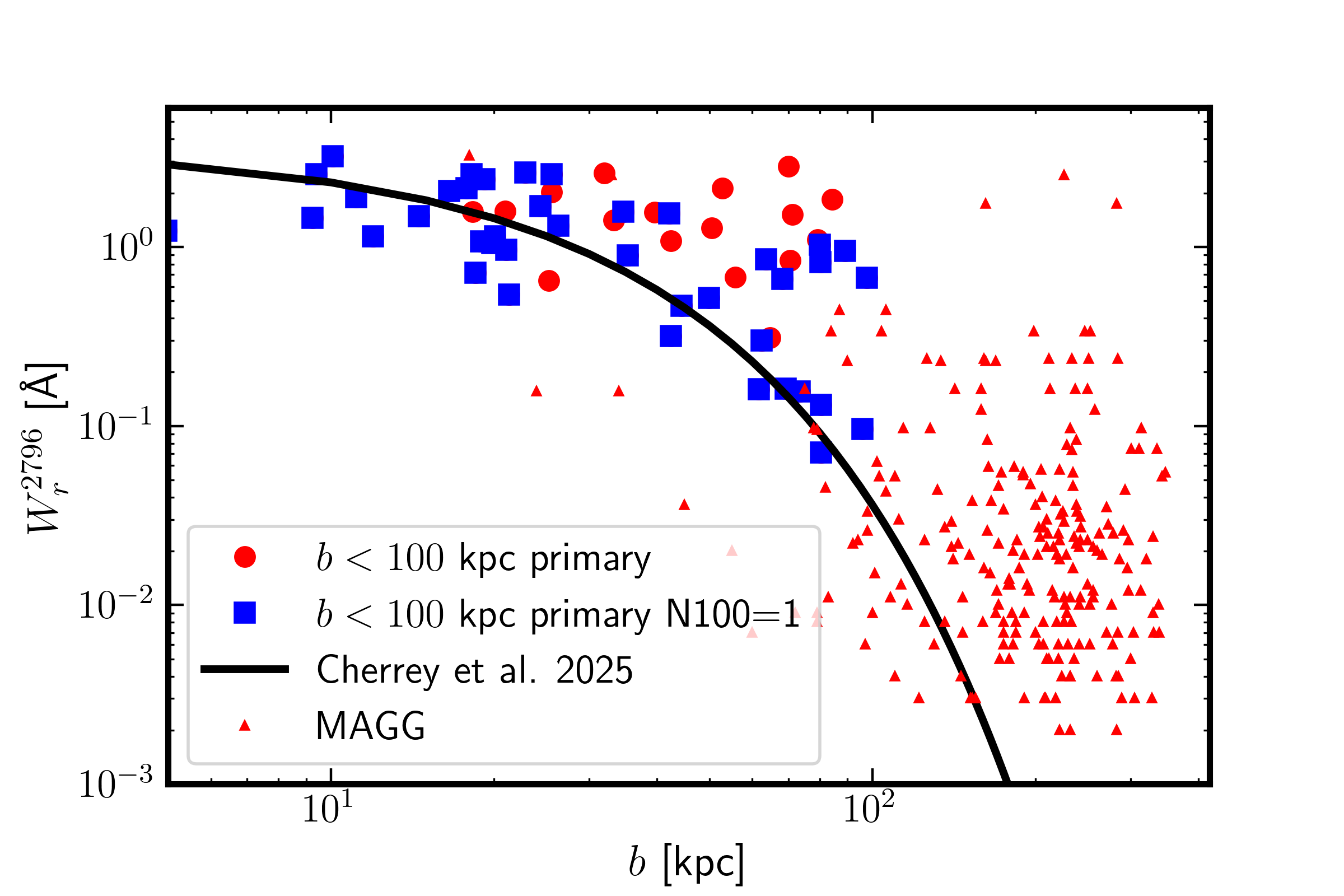}
\caption{REW  ($\Wr$) vs impact parameter ($b$) for {\it primary} absorption-galaxy pairs within 100~kpc.
The solid squares (circles) represent the pairs with only 1 galaxy  (mulitple galaxies) within 100~kpc.
\citet{CherreyM_24a} investigates the $\Wr$-$b$ relation of isolated galaxies  and its dependence with the  physical properties of the host.
The triangles represent the MAGG sample \citep{DuttaR_20a}.
}
\label{fig:rew:b}
\end{figure}

Figure~\ref{fig:rew:b} shows the REW impact parameter ($\Wr-b$) relation. The squares (circles) shows the relation for the {  primary} galaxies with  $N_{100}=1$ ($N_{100}>1$), respectively.
In \citet{CherreyM_23a}, we investigate  the   $\Wr-b$ relation for group-selected pairs and the associated in groups.
In \citet{CherreyM_24a}, we  investigate the $\Wr-b$ relation for isolated galaxies  along with its dependence with respect to SFR, mass, redshift and azimuthal angle $\alpha$.

\subsection{On the effect of pre-selecting sight lines}
\label{section:fcovering}

The MEGAFLOW survey is inherently an { absorption-centered} survey with the pre-selection of sight lines with \MgII\ absorption lines as discussed in Sect. ~\ref{section:megaflow}.
However, the wide wavelength range and, hence, the wide redshift coverage ($0.3<z<1.5$)   allows us to perform a {  galaxy-centered} analysis such as the characterization of the covering fraction; namely the fraction of galaxies  which have a \MgII{} absorption above some column density or REW \citep{SchroetterI_21a} or that of groups comprised of more than five galaxies \citep{CherreyM_23a}.
In \citet{SchroetterI_21a}, we investigated the covering fraction
of \MgII{} absorptions at $1.0<z<1.4$ using a preliminary version of the MEGAFLOW galaxy catalog. 
With the full DR2 catalogue in hand, we aim to revisit  in a separate paper the \MgII{} covering fraction for galaxies across the full redshift range $0.3<z<1.5$, as a function of azimuthal angle, and galaxy properties \citep{CherreyM_24a}.
Here, we address the potential impact of the pre-selection of sight-lines (Sect.~\ref{section:megaflow}).

Such {\it galaxy-centered} analysis is in fact possible in MEGAFLOW for a number of reasons. First,  each MUSE field has $\approx50$ galaxies at $0.3<z<1.5$ observed over 2000 independent channels taking into account the spectral resolution; thus, the pre-selection of fields only affects 5-10\%\ of the sample.  Second, the shape of the REW distribution ${\mathrm d}n/{\mathrm d}W$ in MEGAFLOW is consistent with that of field or random sight-lines 
(as discussed in Sect.~\ref{section:MgII:catalog} and Fig.~\ref{fig:uves:rew}), also   noted in \citet{SchroetterI_21a}. We should note that if the pre-selection of sight lines would bias the results, the galaxy covering fractions
in \citet{SchroetterI_21a} do not seem to be different from other surveys   \citep[e.g.][]{NielsenN_13b,LanT_19a,HuangY_21a,DuttaR_20a}.

Nonetheless, to further quantify the effect of our pre-selection, we performed the following experiment.
We populated MUSE-like fields with galaxies,  each with their own CGM (assumed to be a gas sphere). After applying a MEGAFLOW pre-selection,
we can compute the covering fraction on these selected fields and compare it the covering fraction obtained using all fields (down to the same REW limit).

\begin{figure}
 \centering
 \includegraphics[width=8.5cm]{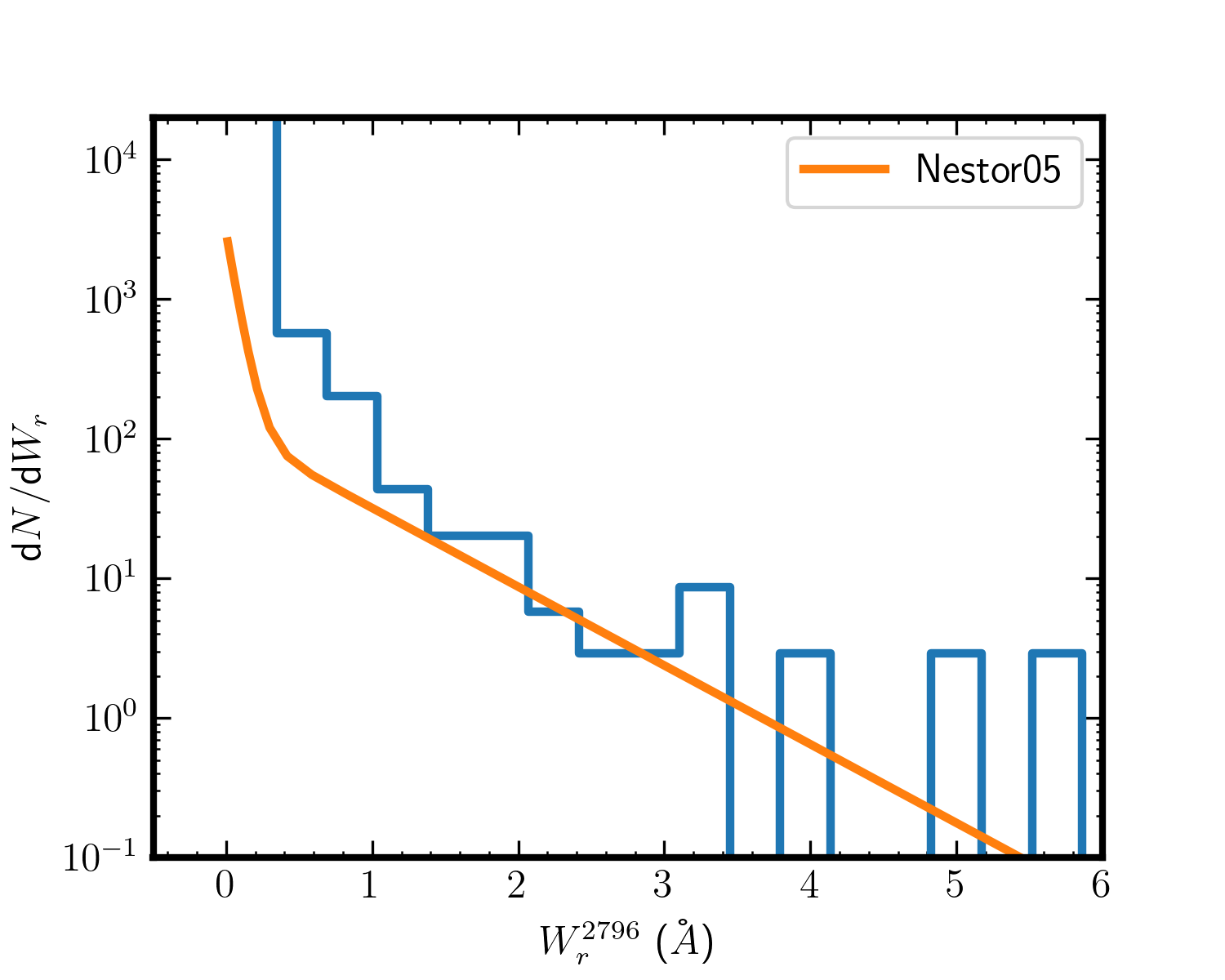}
\caption{Equivalent width distribution \Wr{} for mock galaxies in 50 MUSE-like fields.
The solid line represents the observed distribution of strong \MgII{} absorbers from \citet{NestorD_05a}.
}
\label{fig:mock:REW}
\end{figure}

In particular, we generated 50 MUSE-like fields (500$\times$500 pkpc in projected size corresponding to 1$\times$1 arcmin) and drew 100 galaxies per field from a Poisson distribution across a redshift range ($0.4<z<1.4$).  We assigned a stellar mass from $\log M=9.5$ to $\log M=11.5$ according to a power law distribution of $-1.5$.  We then assigned a truncated sphere of gas around each galaxy \citep[similar to][]{TinkerJ_08a}, with the gas density following a power law  of $\rho\propto r^{-\alpha}$ with $\alpha\approx 2-1.5$, truncated at $R_{\rm tr}=$100~kpc. We set the normalization of the density profile such that the column density reaches $\sim10^{20}$ cm$^{-2}$ at 10~kpc adding a mass dependence ($\propto 0.3 \log M$) following the observed relation \citep{ChenHW_10b}.
 We can then compute the line-of-sight column density for each intercepted sphere and transform it into a REW using the $\Wr-N_{\HI}$ relation from \citet{MenardB_09a}. The resulting equivalent widht $\Wr$ distribution is shown in Fig.~\ref{fig:mock:REW} and is in good agreement with the observed distribution of  strong absorbers \citep{NestorD_05a,ZhuG_13a}.

For each field, we computed the number of absorptions  with a $\Wr$ greater than 0.6\AA, corresponding to pre-selection of strong \MgII{} absorbers. Using only the
$\sim9$-10 fields with 3 or more strong absorptions, mimicking the MEGAFLOW pre-selection, we computed the 2D covering fraction as a function of impact parameter and galaxy mass,  $f_{\rm c}(\log b, \log M)$ using the method used in \citet{SchroetterI_21a} and \citet{CherreyM_23a}.
Using only the selected sight lines, the radius $r_{50}$, radius at which $f_{\rm c}(r_{50}|\log M=10.5)$ is 50\%, is then:
\begin{eqnarray}
\log r_{50}|_{\log M=10.5} &=&1.89\pm 0.34.
\end{eqnarray}
When using all sight lines, this radius is
\begin{eqnarray}
\log r_{50}|_{\log M=10.5} &=&1.91\pm 0.21.
\end{eqnarray}
The mass dependence between \Wr{} and $r_{50}$ is found to be $\propto (0.09\pm0.03)\log M$ for the selected sight lines and $ \propto (0.07\pm0.02)\log M$ for all sight lines.

Apart from the larger statistical uncertainty due to the smaller number of fields, the values are very consistent with each other, and we conclude that there is no inherent bias in covering fractions due to the MEGAFLOW pre-selection criteria.
We refer to \citet{CherreyM_23a} for similar arguments using different assumptions.



\section{Discussion on \OII{} versus continuum selected galaxies}
 \label{section:discussion}

\begin{figure}
\centering
\includegraphics[width=8cm]{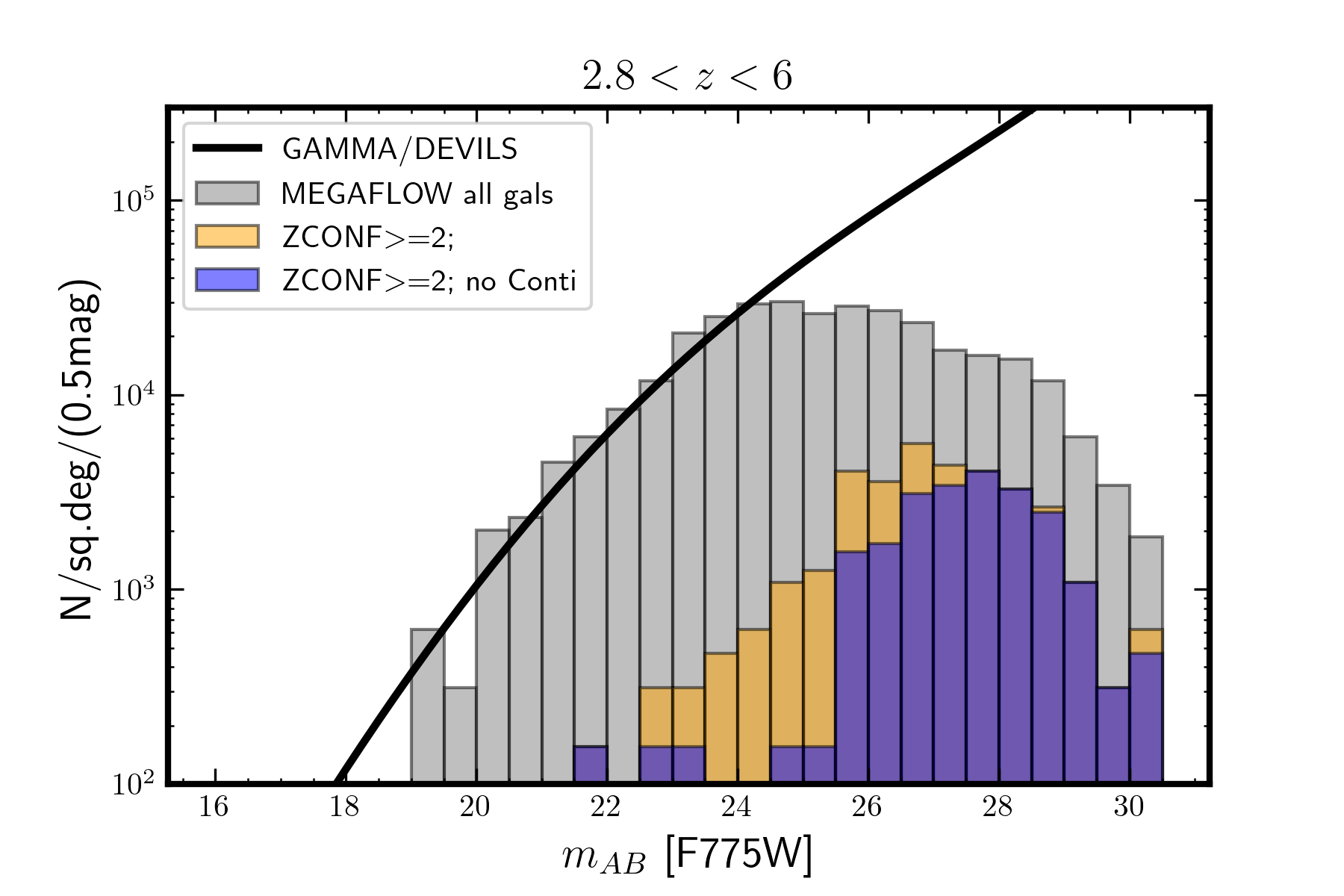}
\includegraphics[width=8cm]{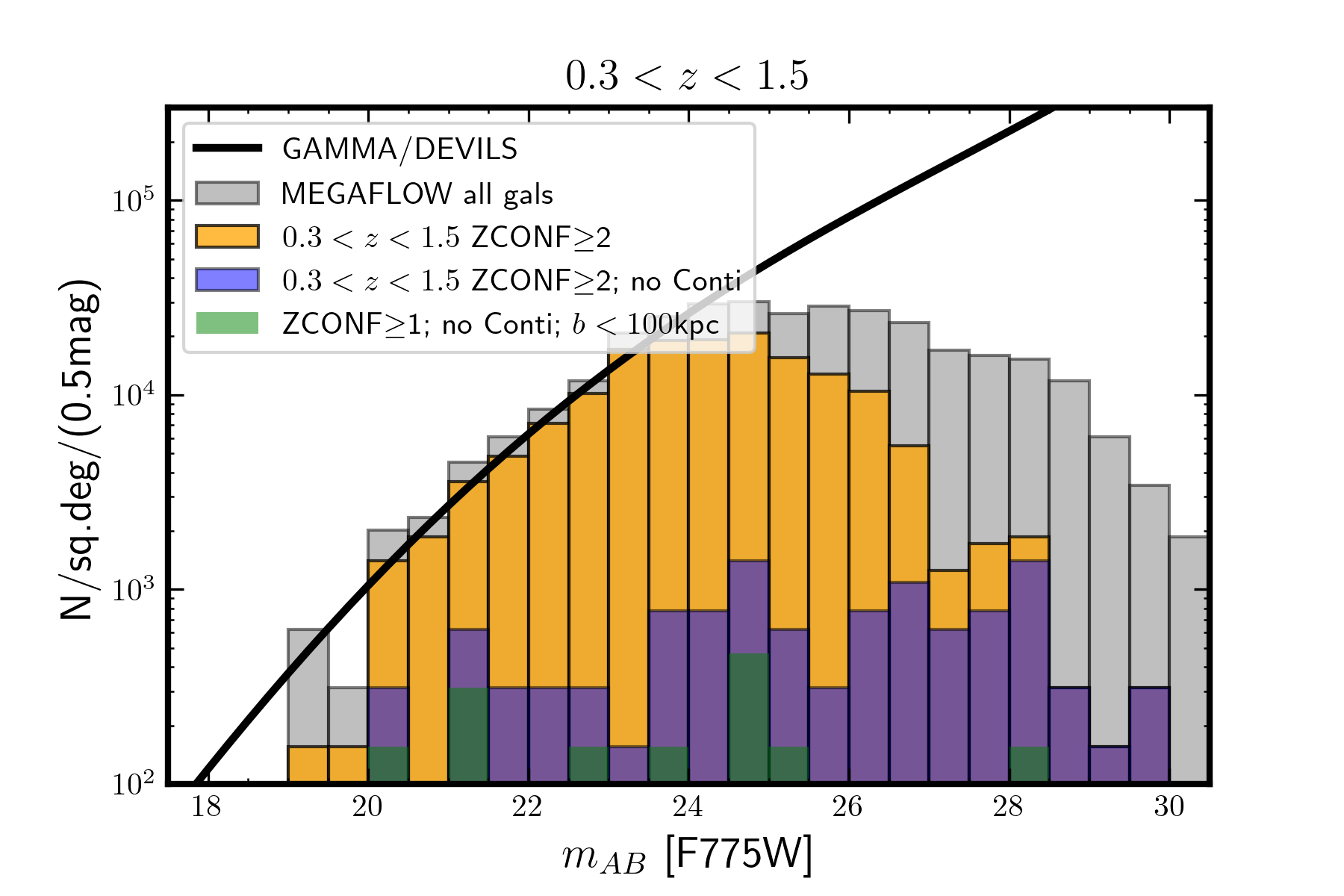}
\caption{
Number counts for galaxies without continuum detection, without a \texttt{WHITE\_ID}. The bottom (top) panel shows the distributions for low (high) redshift galaxies with $0.3<z<1.5$ ($2.8<z<6$) respectively.
Both panels show that there is a significant popultion of   galaxies detected solely on their emission line with \texttt{FELINE}.
The solid line represents the GAMMA/DEVILS survey \citep{devils}.
 }
\label{fig:DR2:histmag}
\end{figure}

\label{section:discussion:dual}

The  MEGAFLOW survey follows a  {  dual} galaxy identification process  based on continuum and on emission lines  (described in Sect.~\ref{section:catalogs:galaxies}) . This  allows for the identification of line  emitters (\OII, \lya) without any continuum in the MUSE observations. 
In the  final catalog (DR2 v2.0) from the  MUSE data, we find that 20\%\ (30\%) of  galaxies with \texttt{ZCONF}$\geq2$ ($\geq 1$) do not have  a continuum ID (\texttt{WHITE\_ID}), respectively.
For \OII{} emitters at $0.3<z<1.5$, the fraction is 8-15\%\ depending on the redshift confidence \texttt{ZCONF} flag  (see Table~\ref{table:nocounterparts}).
For the low-redshift galaxies with \texttt{ZCONF}$\geq2$ that are within 100~kpc and 500~\kms{} of a \MgII{} absorption, 13 out of 138 were detected only from their emission lines.
For the { primary} galaxies  within 100~kpc, 5 out of 58  are detected only from their emission lines. Hence,  10\%\ of  galaxies at $0.3<z<1.5$ have been detected only from their emission lines.

However, not all sources without  {WHITE\_ID} are faint objects below the detection limit.   
Figure~\ref{fig:DR2:histmag} shows the F775W ($\sim i$) magnitude distribution of these galaxies without WHITE\_ID.
At high redshifts ($z>2.8$), the top panel reveals that almost all have  $i=26$---28~mag, namely, they are below the completeness limit.
At low redshifts ($z<1.5$),  the bottom panel reveals a population of galaxies without WHITE\_ID but with bright $i<26$~mag, which can often be traced to failed deblending or to a bright foreground or background object contaminating the flux measurement. The green histogram shows the 13 galaxies with \texttt{ZCONF}$\geq2$ that are within 100~kpc and 500~\kms{} of \MgII{} absorptions, which are very relevant for the MEGAFLOW survey.

\begin{figure*}
\centering
\includegraphics[width=0.8\textwidth]{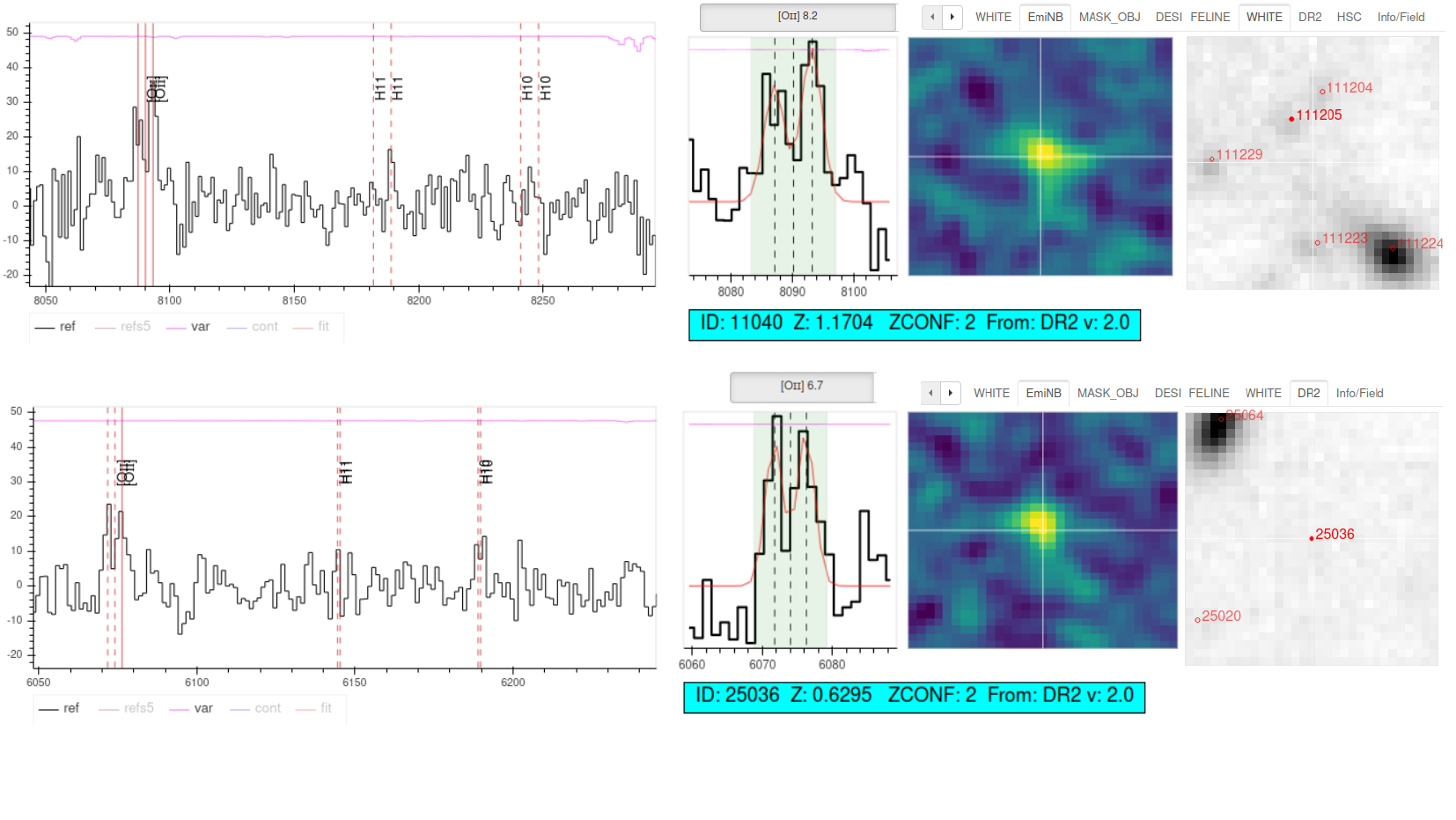}
\caption{Examples of \OII{} emitters without continuum for source ID$=11040$ and ID$=25036$.  More information available on the \texttt{AMUSED} interface. For each source, the MUSE spectra,   \OII{} emission,   pseudo narrow-band image at \OII{}, and the white-light continuum image are shown. 
 }
\label{figure:noconti:examples}
\end{figure*}

The advantage of the dual detection method can be seen in
 Figure \ref{fig:OII_vs_SDSSr}(left), which  shows the $r$-band magnitude versus the \OII\ flux for the { primary} galaxies in MEGAFLOW with or without \MgII{} absorption. From this figure, we can see that the \texttt{FELINE} line detection allows us to detect object beyond the continuum limit of  $r\simeq25$~mag.
Figure~\ref{figure:noconti:examples} shows two examples of galaxies with \OII\ emission and no continuum.

Similarly, in the Muse eXtremely Deep Field [MXDF; 140hr] from \citet{BaconR_23a} who performed also a dual continuum and emission line detection, 
  there are 175 out of 886  \Ly{} emitters (20\%)  that have no counterparts in the HST catalog 
\citep{RafelskiM_15a} and   96 out 886 \Ly{} emitters (10\%)  have no detectable signal in the deep HST/UDF images.
For comparison, the MAGG sample \citep{LofthouseE_20a} is based  on continuum identification and does not include blind emission line galaxies. Similarly, the QSAGE sample \citep{Bielby_2019} is based on continuum identification on shallow 250s WFC3 F140W images. The sample from the MUSE Ultra Deep Field [MUDF; 150hr] of \citet{FossatiM_19a} and \citet{RevalskiM_23a}  used deep (6.5hr) continuum F140W images for their catalog identification.


\begin{table}
\caption{Galaxies without continuum  or without FELINE detection.  
}
\label{table:nocounterparts}
\begin{tabular}{llrrrr}
\hline\hline
ZCONF~\tablefootmark{a} & Redshift~\tablefootmark{b} & $N_{\rm tot}$~\tablefootmark{c} & no cont.~\tablefootmark{d} & no FEL~\tablefootmark{e}\\
\hline
$1,2,3$  & All & 1998 & 730 & 150\\
$2,3$ & All &  1473 & 308 & 90\\
$1,2,3$ & $0.35<z<1.5$ & 1151 & 170 & 73 \\
$2,3$ & $0.35<z<1.5$ & 978 & 84  & 53\\
\hline
\end{tabular}
\tablefoot{
\tablefoottext{a}{Redshift confidence flag \texttt{ZCONF}.}
\tablefoottext{b}{Redshift range.}
\tablefoottext{c}{Number of sources.}
\tablefoottext{d}{Number of galaxies without continuum.}
\tablefoottext{e}{Number of galaxies without \OII.}
}
\end{table}


\section{Conclusions}
\label{section:conclusions}

In this paper, we present the MEGAFLOW survey, a  MUSE and UVES survey designed to better understand the CGM using low-ionization metal \MgII{} lines in 22 QSO fields.
We describe  the survey strategy (Sect.~\ref{section:megaflow}) and detail the MUSE and UVES data  (Sect.~\ref{section:data}) along with the data products  (\ref{section:catalogs}).

The MEGAFLOW survey targeted 22 QSO fields with at least three \MgII{} absorptions lines with rest-frame
$\Wr \gtrsim 0.5$~\AA, leading to 79 \MgII{} absorption lines. From the UVES spectra, we found  additional \MgII{} absorption lines, leading to a total of 127 \MgII{} absorption lines. The rest-frame equivalent width distribution follows the expected power-law distribution of random sight lines, albeit with a boosted normalization (Fig.~\ref{fig:uves:rew})

Regarding this MUSE GTO programme, the observations taken between 2014 and 2019, cover 22 arcmin$^2$, 
two fields (J0014$-$0028, J0937$+$0656) were observed at 10-11hr depths, while the other 20 fields were observed at 2-5hr depths.
The $3\sigma$ sensitivity for emission lines at 7000\AA\ is typically $5\times10^{-19}$~\flux~arcsec$^{-2}$ for the 10hr depth cubes, and 
$7.5\times10^{-19}$~\flux~arcsec$^{-2}$ for the 3-4hr depth cubes. 
As discussed in Sect.~\ref{section:completeness}, the 50\%\ completeness limit for the 10hr (2.5hr) cubes  is \detLimOII{}  (\detLimOIIshallow) \OII\ emitters at $\approx7000$~\AA, and is approximately $i\sim m_{F775W}\approx$\detLimCont{} (\detLimContshallow)  for continuum sources, respectively.

The  final catalog (v2.0) from the MEGAFLOW survey contains 2427 sources, which includes 22 Quasars, 57 stars,  2020 galaxies with \texttt{ZCONF}$\geq1$ and 350 sources with no redshifts (\texttt{ZCONF}$=0$).
As discussed in Sect.~\ref{section:catalogs:galaxies}, we use a dual galaxy identification process based both  on continuum \citep[using \texttt{SExtractor; }][]{BertinE_96a} and on \OII\ emission lines  \citep[using \texttt{FELINE};][]{WendtM_24a}. This approach is similar to the MUSE UDF analysis of \citet{BaconR_23a}.  It has the significant advantage that \OII{} emitters without continuum counterpart can be identified (Fig.~\ref{figure:noconti:examples}).
Overall, we  find that one third (718/1998) of galaxies with redshift flag \texttt{ZCONF}$\geq1$ have no continuum detection down to $r\approx28.5$ m$_{\rm AB}$. For \OII{} emitters at $0.3<z<1.5$,  we find 8 (15) per cent of galaxies without a continuum detection for sources with ZCONF $\geq2$ ($\geq1$) (Table~\ref{table:nocounterparts}).


For strong ($\Wr\gtrsim0.5$~\AA) \MgII{} absorbers in the redshift range $0.35<z<1.5$, where \OII{} can be identified in the MUSE wavelength coverage,  the success rate of detecting the host galaxy within 100~kpc is 90\%{} (Table~\ref{tab:success}). The mean number of galaxies per absorber is $\sim3$ within the MUSE FOV, while 40 (20) absorbers have one (two) galaxies within 100~kpc.

 All but two of the host galaxies have SFR greater than 1~\mpy\ and stellar masses above $10^9$~\msun,
after performing SED fitting on the MUSE data using 13 pseudo-medium filters. Given that this MUSE survey is sensitive to galaxies with $M_\star\gtrsim10^{7.5}$ and SFR $\gtrsim0.01$~\mpy{}
(Fig.~\ref{fig:MS_abs_vs_noabs}), we can rule out the role of LMC-like satellites for strong \MgII{} systems with $\Wr>0.5$~\AA. This  is supported by the fact that we would have  detected  satellites with mass ratio $1:20$ or greater in such cases.

\begin{acknowledgements}
This study is based on observations collected at the European Southern
Observatory under ESO programs listed in Table~\ref{tab:muse_obs} and \ref{tab:uves_obs}.
This work has been carried out thanks to the support of the ANR FOGHAR (ANR-13-BS05-0010), the ANR 3DGasFlows (ANR-17-CE31-0017), and the OCEVU Labex (ANR-11-LABX-0060).  LW acknowledges funding by the European Research Council through ERC-AdG SPECMAP-CGM, GA 101020943.
RB acknowledges support from the ANR L-INTENSE (ANR-CE92-0015).
\end{acknowledgements}

{\it Software:} This  work  makes  use  of  the  following  open  source
software:
\textsc{GalPaK$^{3D}$}  \citep{BoucheN_15a},
\textsc{FELINE} \citep{WendtM_24a},
\textsc{ZAP} \citep{SotoK_16a},
\textsc{MPDAF} \citep{mpdaf},
\textsc{Matplotlib} \citep{matplotlib},
\textsc{NumPy} \citep{numpy},
\textsc{Astropy}  \citep{astropy2018},
\textsc{MAOPPY} \citep{FetickR_19a},
\textsc{Pampelmuse} \citep{KamannS_13a},
\textsc{Photutils} \citep{photutils},
 \textsc{dustmaps}  \citep{GreenG_18a}, and
\textsc{pymc3} \citep{pymc3}.

 \section*{Data availability}
The MUSE and UVES raw data used for this article are available in the ESO archive~\footnote{\url{http://archive.eso.org}.}.
The reduced MUSE data-cubes are available on the \texttt{MuseWise} website~\footnote{\url{https://http://muse-dbview.hpc.rug.nl/QaView}.}
The MUSE and UVES products, the MEGAFLOW catalogs and MUSE advanced data products are  also available on the \texttt{AMUSED} website~\footnote{\url{https://amused.univ-lyon1.fr/project/megaflow}.}. 




\bibliographystyle{aa}



 
\appendix

\section{UVES EW limits}

\begin{figure*}
\centering
\includegraphics[width=0.7\textwidth]{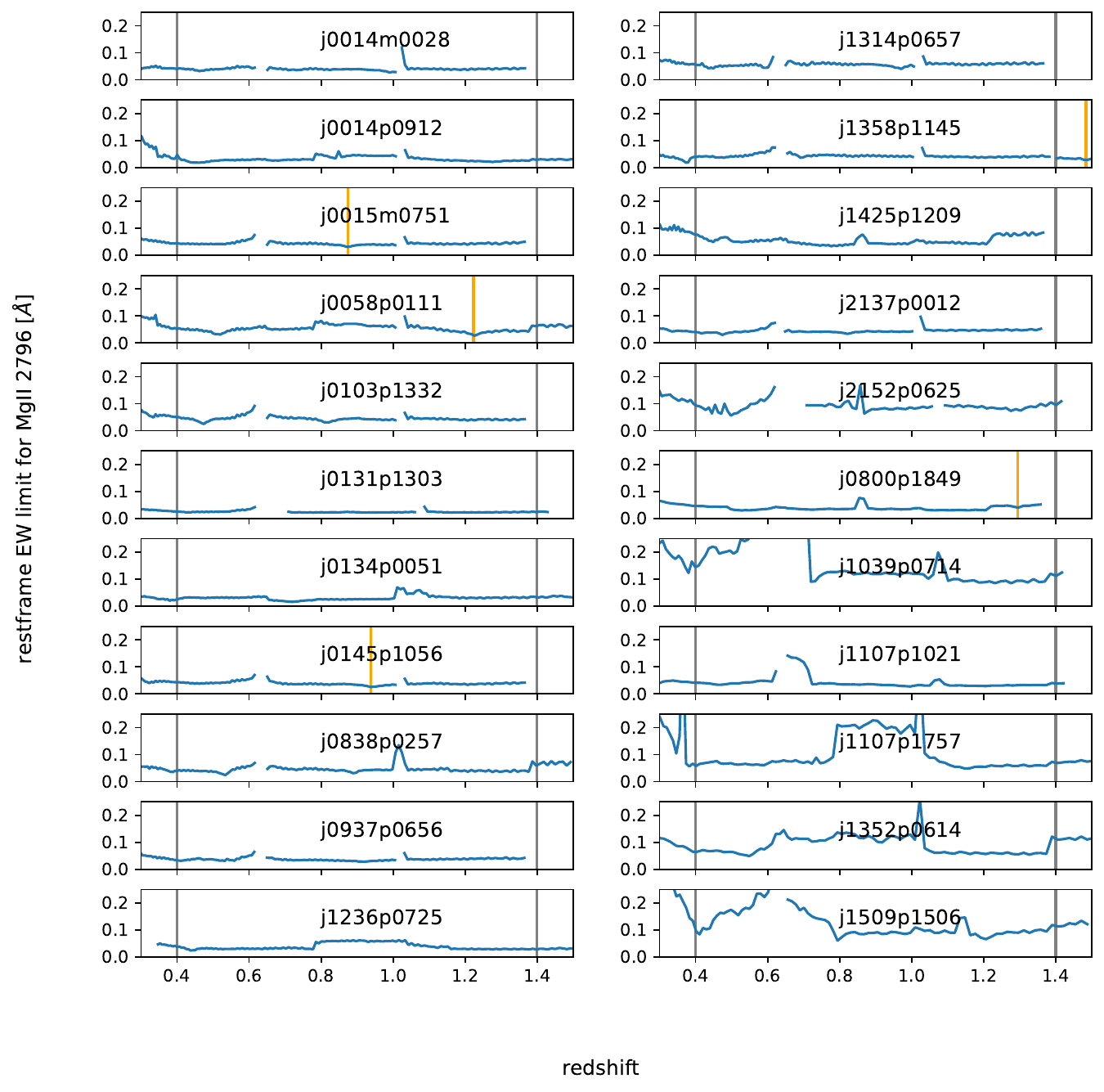}
\caption{Rest-frame \Wr{}  limits (3-$\sigma$) for \MgII{} 2796 per field and redshift. The {\it grey bars} indicate the redshift range of the survey and the {\it orange bars} indicate the QSO's redshift if below z=1.5.}
\label{fig:uves:qsoewlimit}
\end{figure*}

Figure~\ref{fig:uves:qsoewlimit} shows the rest-frame \Wr{} limits (3$\sigma$) for each field used in Fig.~\ref{fig:uves:meanewlimit} as a function of redshifts which ranges from 0.05 to 0.1\AA.

\section{Description of the catalogs and products}
\label{appendix:catalogs}

Table~\ref{table:catalog:dr2} describes the content of the galaxy catalog.
Table~\ref{table:catalog:abs} describes the content of the \MgII\ absorption catalog.
Table~\ref{table:catalog:datasets} describes the available MUSE datasets and associated files.

\begin{figure*}
\centering
\includegraphics[width=15cm]{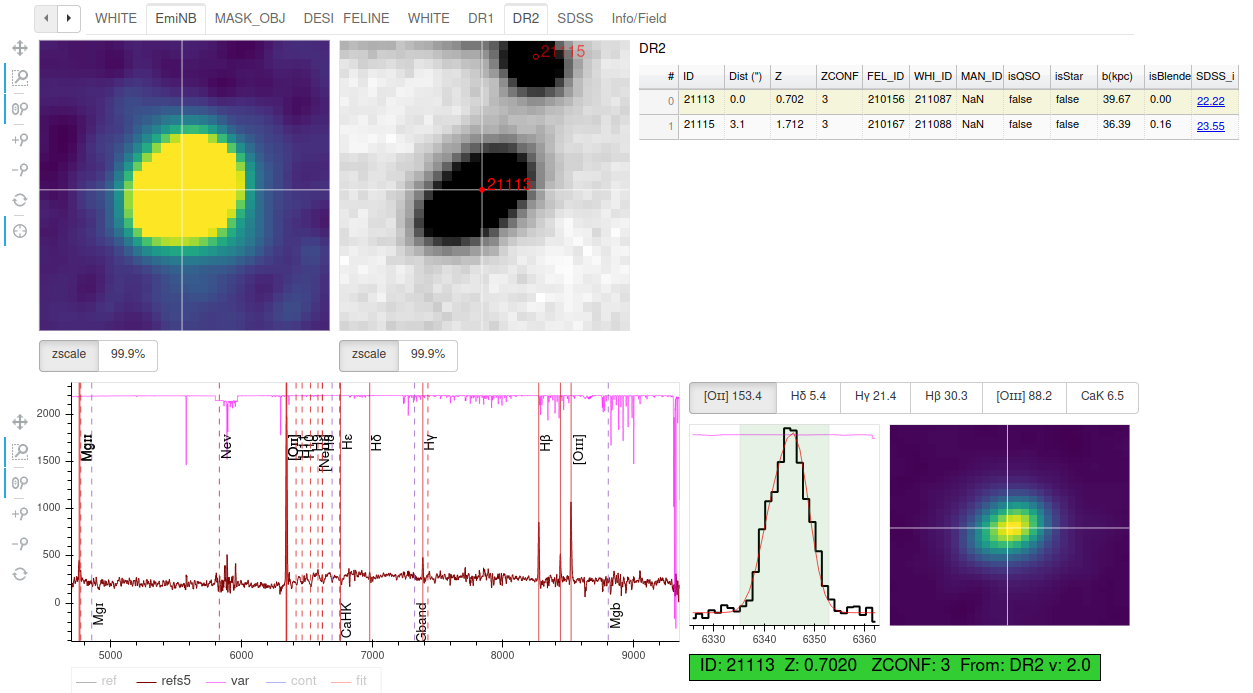}
\caption{Example of a html file summarizing the information regarding each source available from the \texttt{AMUSED} web interface. This source is the $z=0.702$ galaxy ID 21113 with a \MgII\ halo presented in \citet{ZablJ_21a}. }
\label{fig:examples}
\end{figure*}

Figure~\ref{fig:examples} shows an example of an html file  displaying the summary information of each sources, available on the \texttt{AMUSED} interface.

\begin{table*}
\centering
\caption{Column description of the main galaxy catalog.} 
\label{table:catalog:dr2}
\begin{tabular}{lr}
\hline\hline
Col.name & Description \\
\hline
ID &  Unique source identifier \\
FIELD\_NAME & Short name of the field \\
DATASET &  MUSE Dataset (\texttt{beta}, \texttt{psfsub}) \\
FROM &  Source origin (FELINE, MANUAL, WHITE) \\
Z &  Redshift \\
Z\_ERR & Redshift error \\
ZCONF & Redshift Confidence (0,1,2,3)\\
 is\_QSO &  Boolean flag for quasars\\
is\_star & Boolean flag for stars\\
is\_blended & Boolean flag for source blending\\
RA & Right ascension (deg)\\
DEC &  Declination (deg)\\
B\_KPC  & Impact parameter in kpc\\
B\_ARCSEC & Impact parameter in arcsec\\
REFSPEC &   Type of spectrum extraction \\
REFCENTER &  Reference center \\
FELINE\_ID & Source identifier from \texttt{FELINE} (Sect.~\ref{sect:galaxies:feline})\\
WHITE\_ID & Source identifier from \texttt{WHITE} (Sect.~\ref{sect:galaxies:conti})\\
MANUAL\_ID & Source identifier from manual entry\\
SDSS\_r & $r$-magnitude \\
SDSS\_r \_ERR & $r$-magnitude  error \\
SDSS\_i & $i$-magnitude \\
SDSS\_i \_ERR & $i$-magnitude error \\
MAG\_F775W & F775W magnitude \\
MAG\_F775W\_ERR & F775W magnitude error\\
in\_DR1 & Boolean flag for DR1 sources \\
DR1\_id & ID in DR1 \\
\hline
\end{tabular}
\end{table*}

\begin{table*}
\centering
\caption{Column description of the \MgII\ absorption catalog}
\label{table:catalog:abs}
\begin{tabular}{lr}
\hline\hline
Col.name & Description \\
\hline
asb\_id & absorption ID\\
FIELD\_NAME & Short name of the field \\
 z\_abs   & Absorption redshift\\
REW\_2796   & \Wr (\AA)\\
sig\_REW\_2796&    \Wr uncertainty (1$\sigma$) \\
\hline
\end{tabular}
\end{table*}

\begin{table*}
\caption{List of MUSE  datasets and associated files}
\label{table:catalog:datasets}
\begin{tabular}{llll}
\hline\hline
Field name  & Filename& Dataset & Description  \\
\hline
J1122p3344 & \verb|J1122p3344_dr2_zap.expmap_2d_wcs.fits| &   & 2D Exposure map \\
J1122p3344 & \verb|J1122p3344_dr2_zap_wpsf2.corrEBV.fits|   & beta & 3D Data cube~\tablefootmark{a}   \\
J1122p3344 & \verb|J1122p3344_dr2_zap_wpsf2_qsosub.corrEBV.fits| & psfsub &3D Data cube~\tablefootmark{a}  \\
J1122p3344 & \verb|J1122p3344_dr2.corrspec.fits| &   & 1D Applied E(B-V) correction\\
\hline
\end{tabular}
\tablefoot{
\tablefoottext{a}{Filename. The datacubes are in units of $10^{-20}$~\fdlam.}
}
\end{table*}

\end{document}